\documentclass[a4paper,11pt]{article}
\usepackage{jcappub}

\usepackage{aas_macros}
\usepackage{amsmath}
\usepackage{graphicx}
\usepackage{xcolor}
\usepackage{tikz}
\usepackage[caption=false]{subfig}
\usepackage{mathrsfs,mathtools}
\usepackage{physics,amssymb}
\usepackage{bm}
\usepackage{braket}
\usepackage{listings}
\usepackage{cases}
\usepackage{comment}
\usepackage{soul}
\usepackage{cancel}
\usepackage{cases}
\usepackage[utf8]{inputenc}
\usepackage{url}
\usepackage{longtable}
\usepackage[normalem]{ulem}
\usepackage{xspace}
\usepackage{here}
\usepackage{makecell}

\newcommand{\Ax}{\hat{A}_x}
\newcommand{\Ay}{\hat{A}_y}
\newcommand{\Az}{\hat{A}_z}
\newcommand{\Jx}{\hat{J}_x}
\newcommand{\Jy}{\hat{J}_y}
\newcommand{\Jz}{\hat{J}_z}
\newcommand{\Vx}{\hat{V}_x}
\newcommand{\Vy}{\hat{V}_y}
\newcommand{\Vz}{\hat{V}_z}

\newcommand{\bfk}{\mathbf{k}}
\newcommand{\calL}{\mathcal{L}}

\newcommand{\itP}{\mathit{P}}

\newcommand{\hgpc}{h^{-1}{\rm Gpc}}
\newcommand{\hmpc}{h^{-1}{\rm Mpc}}
\newcommand{\gs}{g_*}
\newcommand{\hmsun}{h^{-1}M_\odot}
\newcommand{\cvir}{c_{\rm vir}}

\definecolor{orange}{rgb}{1.0,0.6,0}
\definecolor{DARKBLUE}{HTML}{00008b}
\definecolor{DARKMAGENTA}{HTML}{8b008b}

\begin{document}
\title{Shapes and orientations of massive halos in the statistically anisotropic universe
}

\author[a,b]{Shogo~Masaki,}
\author[b]{Yurino~Mizuguchi,}
\author[c,d]{Shohei~Saga,}
\author[b,d,e]{and~Shuichiro~Yokoyama}

\affiliation[a]{Department of Information Engineering and Institute for Advanced Studies in Artificial Intelligence, Chukyo University, Toyota, Aichi 470-0393, Japan}
\affiliation[b]{Department of Physics, Nagoya University, Nagoya, Aichi 464-8602, Japan}
\affiliation[c]{Institute for Advanced Research, Nagoya University, Nagoya, Aichi 464-8601, Japan}
\affiliation[d]{Kobayashi Maskawa Institute, Nagoya University, Nagoya, Aichi 464-8602, Japan}
\affiliation[e]{Kavli Institute for the Physics and Mathematics of the Universe (WPI), UTIAS, The University of Tokyo, Kashiwa, Chiba 277-8583, Japan}
%\emailAdd{shogo.masaki@gmail.com}
\emailAdd{mizuguchi.yurino.y0@s.mail.nagoya-u.ac.jp}
%\emailAdd{saga@kmi.nagoya-u.ac.jp}
%\emailAdd{shu@kmi.nagoya-u.ac.jp}

\date{\today}

\abstract{
We investigate how statistical anisotropy (SA) in matter distributions affects the distributions of shapes and orientations of cluster-sized halos, using cosmological $N$-body simulations that incorporate SA. 
While the three-dimensional halo shape parameters show little dependence on SA, we find that halo orientations are significantly influenced, with halos tending to align either perpendicular or parallel to the SA direction. 
This SA-induced alignment becomes more prominent for more massive halos.
We also study other vector quantities associated with the dynamics of halos, such as bulk velocity and angular momentum vectors. We find
that their dependences on the SA are smaller than those of the orientation vectors.
Our findings suggest that observational measurements of projected halo shapes derived from galaxy cluster-galaxy lensing could provide a novel probe of SA in the universe.
}

\maketitle
\flushbottom

%%%%%%%%%%%%%%%%%%%
\section{Introduction}
%%%%%%%%%%%%%%%%%%%
Statistical isotropy, also referred to as global isotropy, is a foundational conjecture in cosmology.  
It implies that the statistical properties of density fluctuations are independent of direction.  
While this conjecture is supported by various observations, such as the cosmic microwave background (CMB) and galaxy distributions, theoretical models involving vector fields naturally lead to its violation.  
Thus, the possibility of broken isotropy—known as statistical anisotropy (SA)—in cosmological matter distributions remains open.

SA can arise from anisotropic sources such as vector fields, which could appear in anisotropic inflationary scenarios (see Refs.~\cite{Dimastrogiovanni:2010sm,Soda:2012zm,maleknejad2013gauge} for review), vector dark matter \cite{Hambye:2008bq,Graham:2015rva,Bastero-Gil:2018uel,2025PhRvD.111j3520F}, and vector dark energy models \cite{BeltranJimenez:2008iye}.  
In particular, inflationary models with gauge fields, including the anisotropic inflation model, have been extensively discussed, e.g., in the context of axion cosmology based on the string theory, primordial magnetogenesis, and so on.
Thus, exploring the SA may indirectly provide a means to probe 
such scenarios.

The quadrupolar type of SA, the leading-order term~\cite{Ackerman:2007nb}, 
is characterized by only one magnitude parameter $\gs$, which has been constrained by the CMB measurements of {\it Planck} satellite observations \cite{2013PhRvD..88j1301K,Planck:2018jri,Planck:2019evm,Planck:2019kim}, yielding $|\gs|=\mathcal{O}(10^{-2})$.
Additional efforts have been made to constrain SA using galaxy clustering measurements \cite{pullen2010non,sugiyama18} (see Refs.~\cite{Shiraishi:2016wec,shiraishi21,shiraishi23} for the practical analysis techniques using the polypolar spherical harmonic basis).
Recently, Ref.~\cite{Masaki:2024hzn} performed cosmological $N$-body simulations incorporating quadrupolar SA, and showed that SA induces anisotropic halo bias in the quadrupole moments of galaxy cluster two-point statistics.
Including this newly identified effect in the analysis enables more accurate constraints on $\gs$ 
from galaxy cluster clustering. 

Inspired by the previous work, we, in this paper, investigate how SA affects the shapes and orientations of cluster-sized halos at $z=0$ using cosmological simulations \cite{jing02,2006MNRAS.367.1781A,schneider12}, with the aim of exploring their potential as a new and complementary probe of SA.
Since SA introduces a preferred direction in the initial conditions of matter distributions, and halo formation traces the underlying matter distribution, it is reasonable to expect that halo shapes and orientations may be altered and become somewhat anisotropic in the statistically anisotropic universe.
In fact, observationally, the projected shapes of massive halos have been studied mainly via galaxy cluster-galaxy lensing.
For example, Ref.~\cite{oguri10} measured the distribution of the projected ellipticity of massive halos using galaxy cluster-galaxy lensing signals and found consistency with predictions from cosmological simulations \cite{jing02}.
Ref.~\cite{shin18} obtained the mean ellipticity of galaxy clusters using distributions of member galaxies as well as weak lensing.
If SA affects halo shapes and/or orientations, the observable quantities, such as the projected halo ellipticity, would also exhibit systematic directional features, which could be detected through gravitational lensing observations. In such a case, comparisons between predictions from cosmological simulations with SA and observations of projected ellipticities could serve as a means to constrain $\gs$.
More interestingly, recent analysis of the Dark Energy Survey year-3 shape catalog~\cite{2025arXiv251110005D} has reported a coherent large-scale axial intrinsic alignment signal, which could potentially intersect with SA-related phenomena, raising the possibility that SA might contribute to or be probed through such large-scale alignment signals.

By carefully investigating the halo catalog obtained through the cosmological $N$-body simulations with the SA, we find that the three-dimensional shape parameters \cite{2006MNRAS.367.1781A} show little dependence on the quadrupolar SA parameter $\gs$.
In contrast, halo orientations are more strongly influenced by SA.
Specifically, halos tend to increasingly align perpendicular to the SA-preferred direction for positively larger $\gs$, and parallel for negatively larger $\gs$.
We also find that this alignment effect becomes more pronounced for more massive halos.
In addition to shapes and orientations, we examine the SA effect on vector quantities associated with the dynamics of halos, such as bulk velocities and angular momenta since the SA-distorted matter distributions can affect the bulk velocity fields.
By comparing the SA effects on these halo orientations and dynamical vectors, we conclude that the orientations are more sensitive and serve as a possible new probe of the SA.

The rest of this paper is organized as follows.
In Sec.~\ref{setup}, we describe the cosmological $N$-body simulations incorporating SA, the halo catalogs used in this work, and the measured quantities related to the halo shapes. 
Sec.~\ref{results} presents our results on halo shapes, orientations, and other dynamical vector quantities such as bulk velocities and angular momenta.
We summarize and conclude in Sec.~\ref{conclusion}.

%%%%%%%%%%%%%%%%%%%
\section{Simulations}\label{setup}
%%%%%%%%%%%%%%%%%%%

We first outline the formulations used to incorporate the SA into cosmological $N$-body simulations.
We then describe the simulation settings and define the ellipsoidal parameters that characterize the shapes of halos identified in the SA simulations.

%%%%%%%%%%
\subsection{Cosmological $N$-body simulations with SA}
%%%%%%%%%%
We follow Ref.~\cite{Masaki:2024hzn} to perform cosmological $N$-body simulations with SA.
First, let us introduce the SA in matter distributions.
In our study, we consider the quadrupolar type of SA.
The power spectrum of the initial linear matter overdensity field is written by $\left\langle \delta_\mathrm{m}(\bfk_1)\delta_\mathrm{m}(\bfk_2)\right\rangle =(2\pi)^3 \delta^{(3)}(\bfk_1+\bfk_2)\itP_\mathrm{m}(\bfk_1)$ with
\begin{equation}
    \begin{aligned}
        \itP_\mathrm{m} (\bfk)=\left\lbrack 1+\frac{2}{3}\gs\calL_ 2(\mu)\right\rbrack\bar{P}_\mathrm{m}(k),
        \label{eq:pksa}
    \end{aligned}
\end{equation}
where $\gs$ describes the magnitude of SA, $\bar{P}_\mathrm{m}(k)$ means the isotoropic component of the matter power spectrum, $\calL_2 := \frac{1}{2}(3\mu^2 -1)$ is the second-order Legendre polynomial with $\mu := \hat{\mathbf{k}}\cdot \hat{\mathbf{d}}$ and $\hat{\mathbf{k}}:= \bfk/|\bfk|$, and $\hat{\mathbf{d}}$ is the directions related to SA.

SA in matter distributions is incorporated using Eq.~(\ref{eq:pksa}) when generating initial conditions (ICs), where the direction of SA is set to $\hat{\mathbf{d}}=(0,~0,~1)$.
We use the Boltzmann code {\sc CAMB} \cite{camb} to compute the isotropic linear matter power spectrum $\bar{P}_\mathrm{m}(k)$ in Eq.~(\ref{eq:pksa}) at the initial redshift $z_{\rm ini}=31$.
Based on the calculated $\bar{P}_\mathrm{m}(k)$, ICs are generated using the second-order Lagrangian perturbation theory  \cite{scoccimarro98,Crocce06a,nishimichi09}.
We input the ICs into the cosmological $N$-body solver {\sc Gadget-2} \cite{gadget,gadget2} to follow the evolution of matter distributions in statistically anisotropic universes.
\begin{table}[ht]
\caption{Summary of the simulation settings:
$N_{\rm part}$, $L_{\rm box}$, $m_{\rm part}$, and $\gs$ stand for the number of simulation particles, the size of the simulation box, the mass of a simulation particle, and the magnitude of the quadrupolar-type SA, respectively.}
\centering
\vspace{1mm}
\begin{tabular}{ccccc}
\hline
run & $N_{\rm part}$ & \makecell{$L_{\rm box}$\\$[\hgpc]$} & \makecell{$m_{\rm part}$\\$[\hmsun]$} & $\gs$ \\
\hline
L05 & $512^3$ & $0.5$ & $8.16\times10^{10}$ & $0,~\pm0.1,~\pm1$ \\
\hline
L2 & $1024^3$ & $2$ & $6.53\times10^{11}$ & $0,~\pm0.1,~\pm0.5,~\pm1$ \\
\hline
L4  & $1024^3$ &$4$ & $5.22\times10^{12}$ & $0,~\pm0.1,~\pm1$ \\
\hline
\end{tabular}
\label{table: simulations}
\end{table}
Except for assuming Eq.~\eqref{eq:pksa} as the initial matter power spectrum, in the $N$-body simulation, we assume the standard flat-$\Lambda$ cold dark matter cosmology with the {\it Planck}'s best fit cosmological parameters as $\Omega_{\rm m0}=0.3156,~\Omega_{\Lambda0}=0.6844,~H_0=100h=67.27~{\rm km~s^{-1}~{\rm Mpc}^{-1}},~n_{\rm s}=0.9645$ and $A_{\rm s} = 2.2065\times10^{-9}$ \cite{planck-collaboration:2015fj}.
We perform three simulations characterized by the number of particles $N_{\rm part}$ and the box size $L_{\rm box}$, as summarized in Table~\ref{table: simulations}.
To study the $\gs$-dependence of the halo properties, for each simulation we vary the $\gs$ values as listed in the table.
For each simulation setting, we conducted three realizations, resulting in a total of 51 realizations.

\begin{figure*}[htbp]
        \centering
        \includegraphics[width=\textwidth]{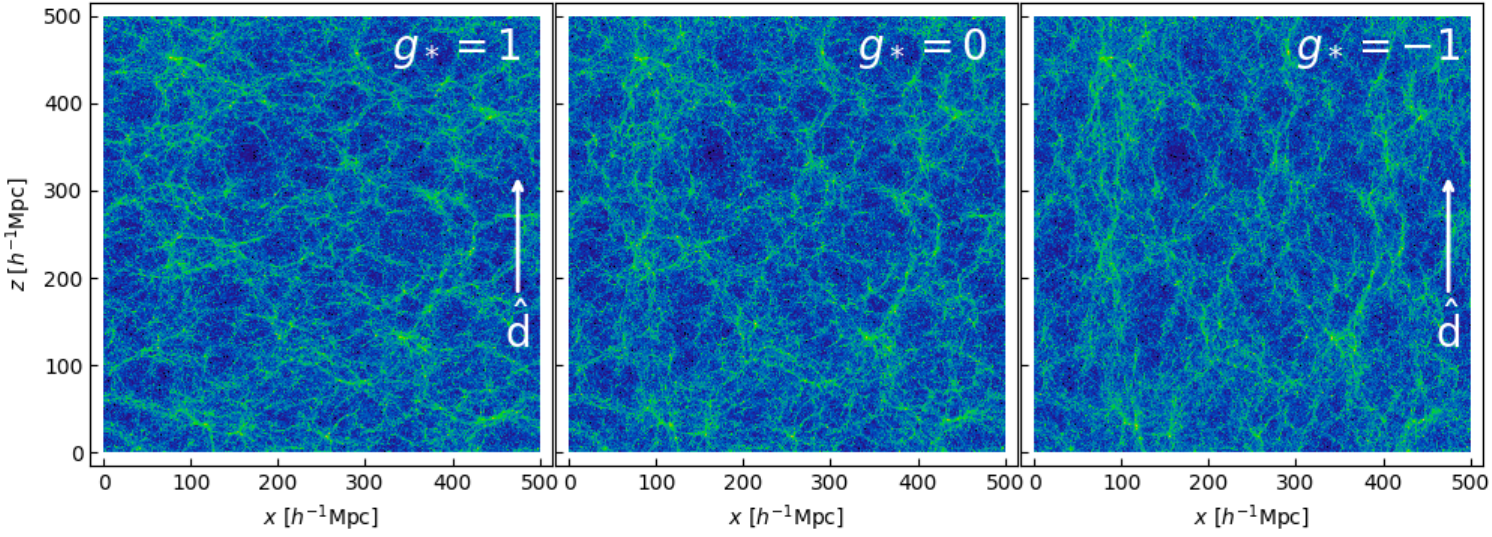}
    \caption{Visualizations of the matter distributions at the redshift $z=0$ in the $x$-$z$ plane,
    in the simulated universes with $\gs= +1$ (left), $0$ (middle), and $-1$ (right).
    Each panel shows a $50~\hmpc$-thick slice along the $y$-direction taken
from the realization with the same initial random seed in the L05 run. The SA direction $\hat{\mathbf{d}}$ is shown in each panel.}
    \label{fig:denmap}
\end{figure*}
To facilitate a visual understanding of the impact of $\gs$, Fig.~\ref{fig:denmap} shows the matter distributions at $z=0$ in the $x$-$z$ plane for the simulated universes with $\gs= +1$ (left), $0$ (middle), and $-1$ (right).
Each panel describes a $50~\hmpc$ thick slice along the $y$-direction taken
from the realization with the same initial random seed in the L05 run.
Compared to the middle panel with $\gs=0$ (i.e., the isotropic case), the matter distribution is elongated perpendicular to the SA direction $\hat{\mathbf{d}}$ for $\gs=+1$ (left) and along $\hat{\mathbf{d}}$ for $\gs=-1$ (right).

%%%%%%%%%%
\subsection{Halos as ellipsoids}
%%%%%%%%%%
To identify halos in the simulated matter distributions,
we use the halo finder code {\sc Rockstar} \cite{Behroozi:2013}. 
We define the halo mass as the mass enclosed within a region whose mean density is 200 times the background matter density, denoted as $M_{\rm 200b}$.
Since measuring halo shapes requires a sufficient number of particles, we analyze only halos that contain more than approximately 200 particles.
On the other hand, a larger simulation box is required to obtain more massive halos due to their lower abundance.
Taking these factors into account, we use halos with masses of $\log_{10}[M_{\rm 200b}/(\hmsun)] \in (13,~13.5]$,~$(14,~14.5]$, and $(15,~15.5]$ from the L05, L2, and L4 runs, respectively.\footnote{Our analyses do not include subhalos.}

To describe the shapes of halos, we regard the halos as ellipsoids characterized by three axes, $a$, $b$, and $c$, with $a \geq b \geq c$.
Based on these three axes, we introduce normalized variables that characterize the halo shapes \cite{franx91,2006MNRAS.367.1781A}
\begin{align}
s:=\frac{c}{a},~q:=\frac{b}{a},~T:=\frac{a^2 - b^2}{a^2 - c^2}=\frac{1-q^2}{1-s^2},
\end{align}
where $s$ and $q$ are the smallest- and intermediate-to-largest axis ratios, respectively, and $T$ is the so-called triaxiality of halos.
For these shape parameters, we directly use the values of $s$ and $q$ produced by {\sc Rockstar}, which are measured using the weighted inertia tensor of halos \cite{2006MNRAS.367.1781A}.

As for the parameter characterizing the orientations of halos, we utilize the vector assigned for each halo in the {\sc Rockstar} halo catalog, $\mathbf{A}=\left(A_x,~A_y,~A_z\right)$, which corresponds to the orientation of the largest axis of the ellipsoid times the halo size.
To clearly show the $\gs$-dependence of the orientations of halos, we use the arranged vector with respect to the $z$-direction (i.e., the SA direction in this work) as
\begin{equation}\label{eq.defA}
\begin{aligned}
    &\mathbf{\hat A}=\left(\Ax,~\Ay,~\Az\right)\\
    &=
    \begin{cases}
        \left(A_x,~A_y,~A_z\right)/|\mathbf{A}| & \text{for $A_z\ge 0$}\\
        \left(-A_x,~-A_y,~-A_z\right)/|\mathbf{A}|& \text{for $A_z < 0$},
    \end{cases}
\end{aligned}
\end{equation}
so that the probability distribution functions (PDFs) of the orientations of halos are uniformly random in the isotropic universe.
The normalization is performed to enable comparison of results across different halo mass ranges.

%%%%%%%%%%%%%%%%%%%
\section{Results}\label{results}
%%%%%%%%%%%%%%%%%%%
In this section, we present
our results on the shapes and orientations of halos at $z=0$ in the statistically anisotropic universe.

%%%%%%%%%%
\subsection{Halo shapes} \label{the halo shapes}
%%%%%%%%%%
%
\begin{figure*}[htbp]
        \centering
        \begin{minipage}{0.45\textwidth}
        \centering
        \includegraphics[width=\columnwidth]{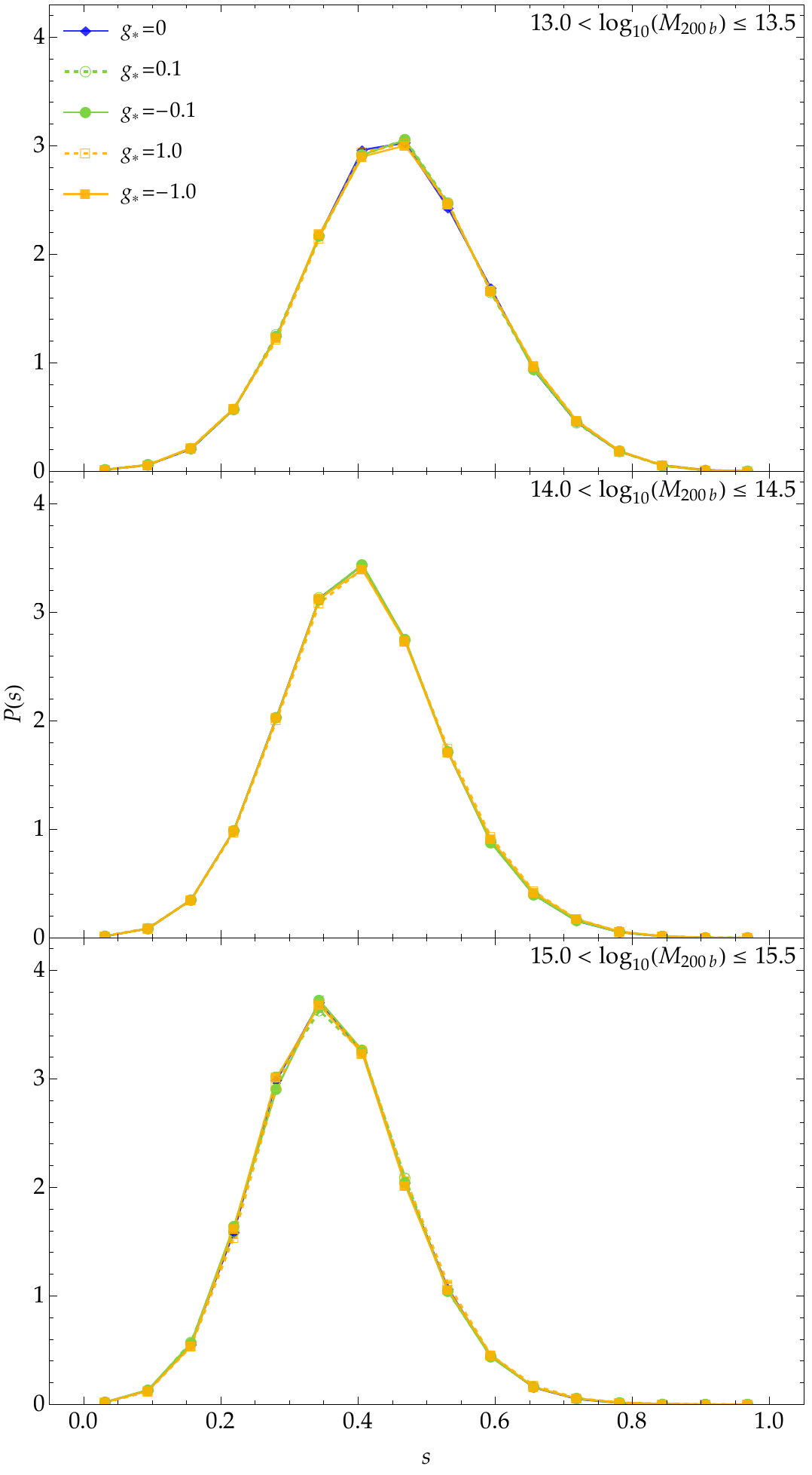}
        \end{minipage}
         \centering
         \begin{minipage}{0.45\textwidth}
         \centering
        \includegraphics[width=\columnwidth]{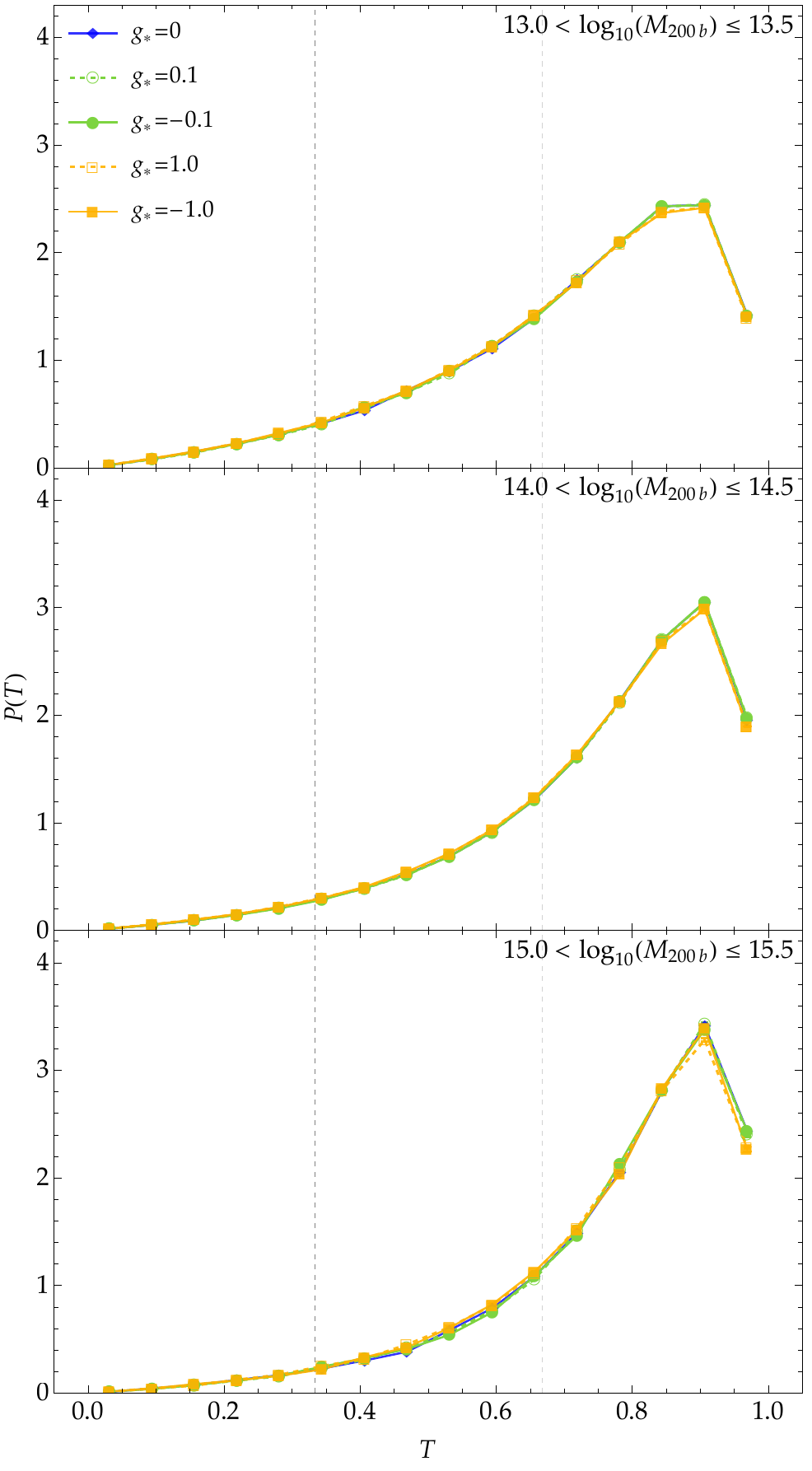}
        \end{minipage}
    \caption{Left: PDFs of $s$, $P(s)$, for the three halo mass ranges of $\log_{10}[M_{\rm 200b}/(\hmsun)]\in(13,~13.5],~(14,~14.5]$ and $(15,~15.5]$ obtained from the L05, L2, and L4 runs, respectively.
    Each color line shows the result for $\gs =0$ (blue),~$\pm 0.1$ (green), and $\pm 1.0$ (orange), and the dashed (solid) lines are for the cases with the positive (negative) sign.
    We note that all five curves are almost indistinguishable. Right: Same as the left panel but for the PDFs of $T$, $P(T)$.
    The vertical dashed line in each panel indicates the boundaries between halo shape categories: ``oblate'' for $T < 1/3$, ``triaxial'' for $1/3 < T < 2/3$, and ``prolate'' for $2/3 < T < 1$. Similar to the left panel, all five curves almost overlap.
    }
    \label{fig:ST_prob}
\end{figure*}
We investigate the effect of SA on the shapes of halos, and in particular, we focus on the parameters $s$ and $T$.

The left panel of Fig.~\ref{fig:ST_prob} shows the $\gs$-dependence of the PDFs of $s$, $P(s)$, for the three halo mass ranges of $\log_{10}[M_{\rm 200b}/(\hmsun)]\in(13,~13.5],~(14,~14.5]$ and $(15,~15.5]$ obtained from the L05, L2, and L4 runs, respectively.
Each color line in this figure shows the result for $\gs =0$ (blue),~$\pm 0.1$ (green), and $\pm 1.0$ (orange), with dashed and solid lines indicating the positive and negative signs, respectively.
The error bars represent the standard error across the three realizations for each value of $\gs$ within a given run (i.e., fixed $L_{\rm box}$ and $N_{\rm part}$).
We note that all the PDFs in this paper are normalized to integrate into one over the entire domain.
Though we focus on the relatively massive halos, the shape of the PDF itself seems to be
consistent with the previous studies in the statistically isotropic universe (see, e.g., Ref.~\cite{2006MNRAS.367.1781A}), e.g., the peak of $P(s)$ is around $s=0.4$, and a weak mass dependence of the peak position is also visible in this figure.

Regarding the $\gs$-dependence, Fig.~\ref{fig:ST_prob} 
shows that even for values of $\gs$ much larger than those currently allowed by the CMB observations, the distributions are still difficult to distinguish from the isotropic case.
As shown later in Sec.~\ref{the halo orientations} (Fig.~\ref{fig:Conditional_PDF_Az_s}), the conditional PDFs $P(\hat{A}_z \mid s)$ exhibit a dependence on $\gs$.
We discuss why this dependence is largely suppressed after marginalization.
%As for the $\gs$-dependence, from this figure, even in the statistically anisotropic universe with the $\gs$ values much larger than %those currently favored by the CMB observations, it is hard to find the difference from the isotropic one.
We also observe a similarly negligible deviation from the isotropic case in the PDFs of the triaxiality, 
 $P(T)$, shown in the right panel of Fig.~\ref{fig:ST_prob}.
Thus, we conclude that for the cluster-sized massive halos, the SA effect on the shape is negligible. 
% \revsout{We expect that the above results on the shapes can be understood as follows. 
% As discussed in Refs.~\cite{shiraishi23, Masaki:2024hzn}, the power spectrum of initial linear matter overdensity fields given by Eq.~\eqref{eq:pksa} can be rewritten as
% $
%          \itP_{\mathrm{m}}(\bfk)=\left\lbrack1+\calG_{ij}\hat{k}_i\hat{k}_j\right\rbrack\bar{P}_\mathrm{m}(k)~, \label{eq:pk_gij}
% $
% with a global traceless tensor field:
% $\calG_{ij} := \gs\left(\hat{d}_{i}\hat{d}_{j}-\frac{1}{3}\delta_{ij}\right).$
% This means that, in our setup, the SA can be understood as the coupling between the matter density field and the global tensor field, and hence the SA effect is considered to be global.
% Thus, such a global effect is not expected to have much impact on the halo shape, which would highly depend on nonlinear dynamics in the halo-sized local region.}

%%%%%%%%%%
\subsection{Halo orientations}\label{the halo orientations}
%%%%%%%%%%
%
\begin{figure*}
  \centering
  \begin{minipage}{0.45\textwidth}
    \centering
    \includegraphics[width=1.05\columnwidth]{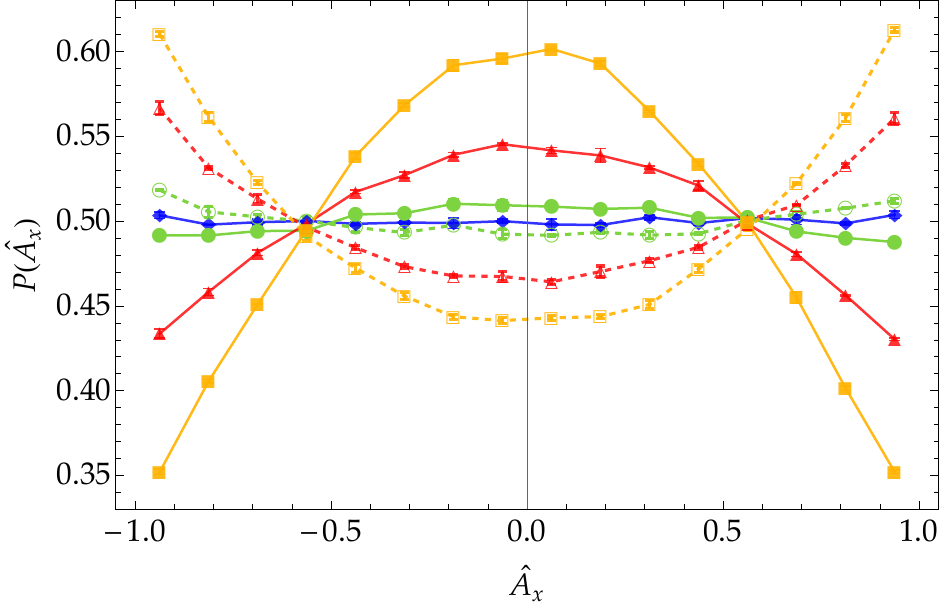}
  \end{minipage}
  \hspace{4mm}
  \begin{minipage}{0.45\textwidth}
    \centering
    \includegraphics[width=1.05\columnwidth]{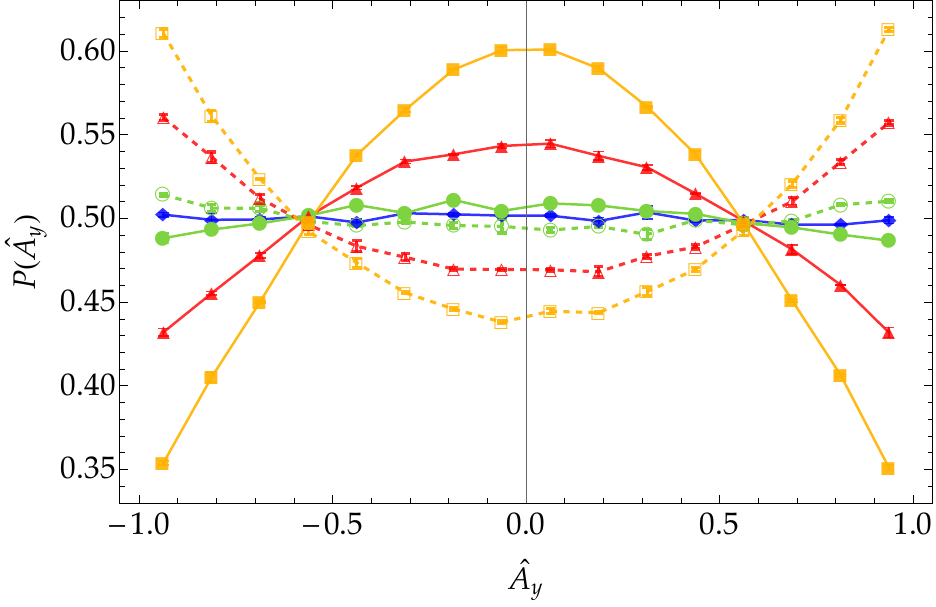}
  \end{minipage}
  \vspace{3mm}
  
  \begin{minipage}{0.4225\textwidth}
    \centering
    \includegraphics[width=1.1\columnwidth]{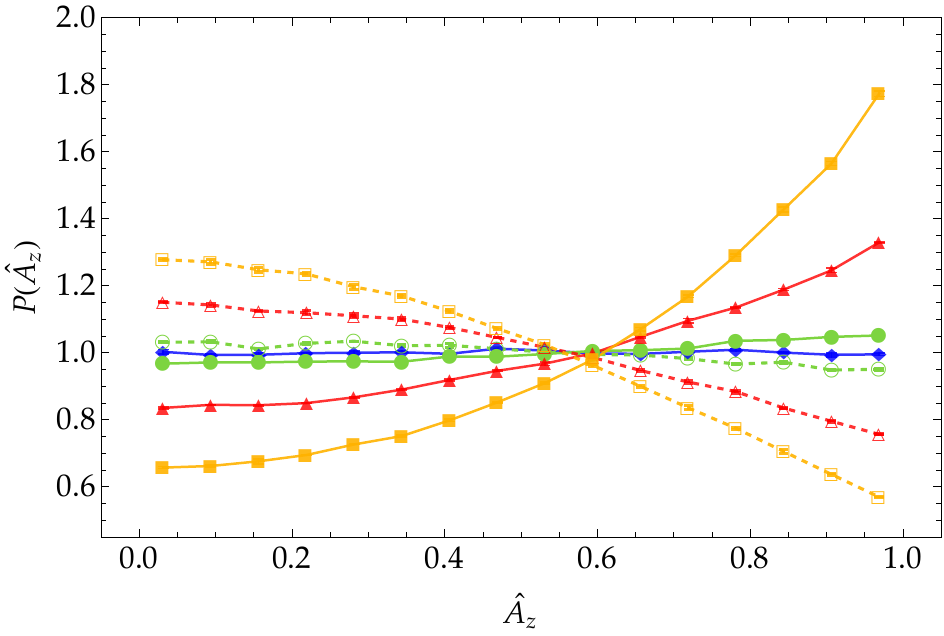}
  \end{minipage}
  \hspace{15mm}
  \begin{minipage}{0.4\textwidth}
    \centering
    \includegraphics[width=1.05\columnwidth]{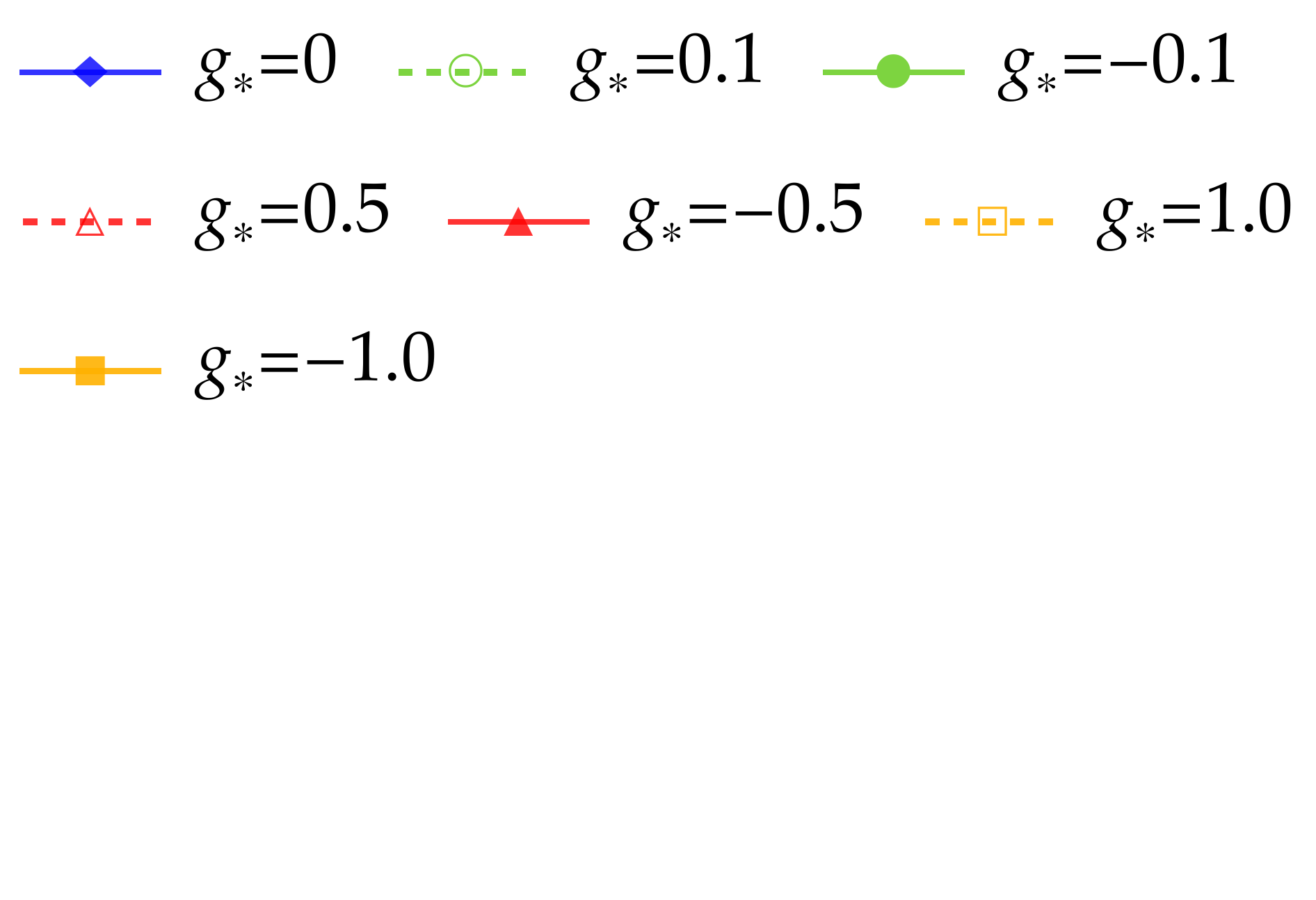}
  \end{minipage}
  \caption{PDFs of $\Ax,~\Ay$ and $\Az$, i.e., the halo orientations, for the halos with masses of $14<\log_{10}[M_{\rm 200b}/(\hmsun)]\leq14.5$ from the L2 run.
  Each color line shows the result for $\gs =0$ (blue),~$\pm 0.1$ (green),~$\pm 0.5$ (red), and $\pm 1.0$ (orange), and the dashed (solid) lines are for the cases with the positive (negative) sign.}
  \label{fig:A_all}
\end{figure*}

As we discussed in the previous subsection, we found that the effect of SA on the orientation-marginalized PDFs of the halo shape parameters %the halo shape
is negligible.
To explore whether there are any halo properties that are affected by the SA, we next examine the orientation of the halo as an ellipsoid.
In Fig.~\ref{fig:A_all}, we show the PDFs of the components of the vector $\mathbf{\hat A}$, which characterizes the orientation of the largest axis of each halo.
Here we use the halos with masses in the range of $14<\log_{10}[M_{\rm 200b}/(\hmsun)]\leq14.5$, taken from the L2 run only. 
Each color line shows the result for $\gs =0$ (blue),~$\pm 0.1$ (green),~$\pm 0.5$ (red), and $\pm 1.0$ (orange), and the dashed (solid) lines are for the cases with the positive (negative) sign.
Note that $\Ax$ and $\Ay$ take values in the range $[-1,~1]$, while $\Az$ takes values in the range $[0,~1]$ due to our setup described in Eq.~(\ref{eq.defA}).

In the isotropic case, i.e., $\gs = 0$, $P(\Ax),~P(\Ay)$ and $P(\Az)$ are completely flat, indicating that the distribution of the orientations of halos is uniformly random.\footnote{This does not imply that the spatial correlations of the halo orientations are zero, as discussed in the context of the intrinsic alignment of galaxies \cite{croft00,heavens00,lee00}.}
Unlike the shape parameters shown in Fig.~\ref{fig:ST_prob}, the halo orientations exhibit a clear dependence on $\gs$.
As $\gs$ increases positively, both $P(\Ax)$ and $P(\Ay)$ exhibit an increasing curvature, developing deeper minima at $\Ax = 0$ and $\Ay = 0$, and more pronounced maxima at $\Ax = \pm 1$ and $\Ay = \pm 1$, respectively.
In contrast, $P(\Az)$ shows the opposite trend, peaking at $\Az=0$ and reaching a minimum at $\Az=1$.
These behaviors indicate that, for larger positive values of $\gs$, halos as ellipsoids tend to align parallel to either the $x$-axis or the $y$-axis.
As $\gs$ decreases (i.e., becomes more negative), $P(\Az)$ becomes a steeper increasing function of $\Az$, showing a higher peak at $\Az=1$ and a lower minimum at $\Az=0$.
Similarly, $P(\Ax)$ and $P(\Ay)$ show the opposite behavior, peaking at $\Ax=0$ and $\Ay=0$ and reaching minima at $\Ax=\pm1$ and $\Ay=\pm1$, respectively.
This means that, for negative values of $\gs$, halos tend to align parallel to the $z$-axis.

In summary, halo orientations tend to increasingly align perpendicular to $\hat{\mathbf{d}}$ for larger positive values of $\gs$, and parallel to $\hat{\mathbf{d}}$ for lower negative values.
This trend is consistent with the visualizations of matter distributions on large scales presented in Fig.~\ref{fig:denmap}.
As discussed in the previous subsection, the SA in our setup can be considered as the effect of the global tensor field, and then the alignment of the halos can be affected by the global structure (see, e.g., Ref.~\cite{shiraishi23}).

Let us mention a few behaviors of the PDFs observed in Fig.~\ref{fig:A_all} that are related to the symmetry in our setup.
First, $P(\Ax)$ and $P(\Ay)$ are nearly identical.
This is because the $x$- and $y$-axes are both perpendicular to the SA direction $\hat{\mathbf{d}} = (0,~0,~1)$, and there is no statistical distinction between the $x$- and $y$-directions.
Second, $P(\Ax),~P(\Ay)$, and $P(\Az)$ are not symmetric with respect to the sign of $\gs$, e.g., between $\gs = +1$ and $\gs = -1$. As discussed above, in the case of $\hat{\mathbf{d}}=(0,~0,~1)$, halos align in parallel to the $x$- or $y$-axes for positive values of $\gs$ while the $z$-axis for negative values.
Consequently, the maximum amplitudes of $|P(\Ax)-0.5|,~|P(\Ay)-0.5|$ and $|P(\Az)-1|$ are not equal between positive and negative $\gs$; they are suppressed for positive $\gs$ because the halo orientations are split between two perpendicular axes.

\begin{figure*}[htbp]
        \centering
        \begin{minipage}{0.45\textwidth}
        \centering
        \includegraphics[width=\linewidth]{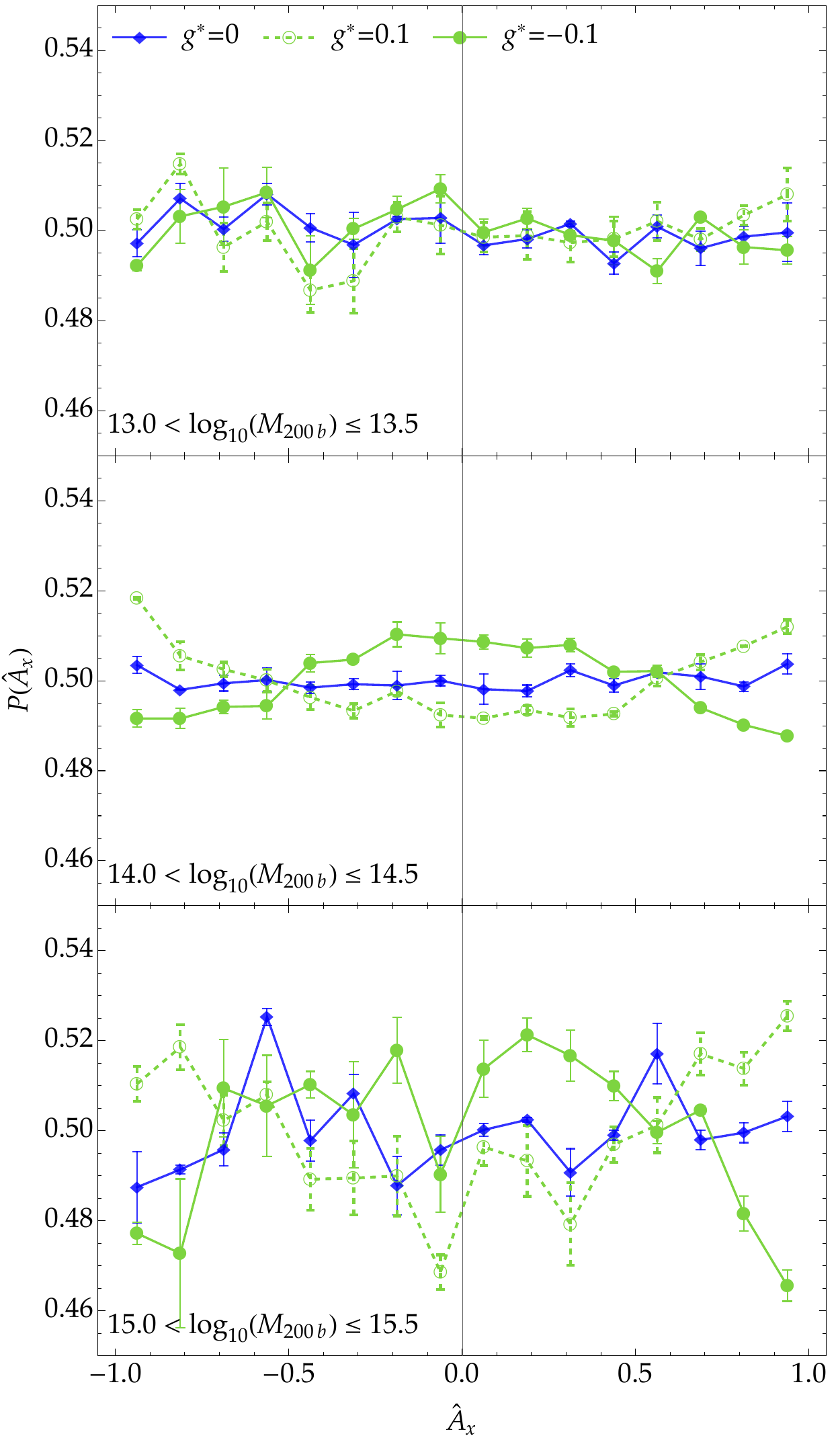}
        \end{minipage}
        \centering
        \begin{minipage}{0.45\textwidth}
        \centering
        \includegraphics[width=\linewidth]{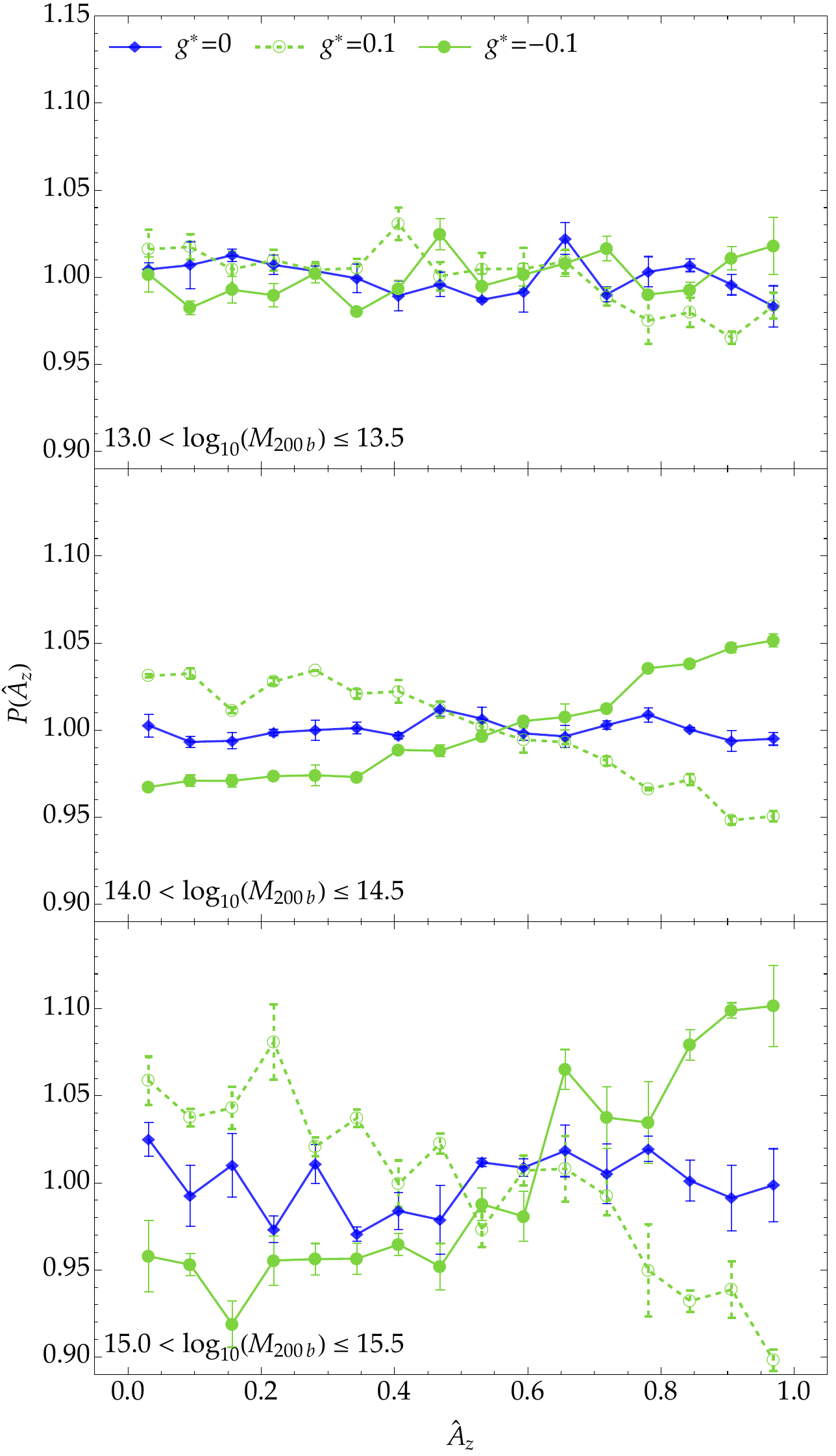}
        \end{minipage}
        \caption{Left: PDFs of $\Ax$, $P(\Ax)$, for the halo mass ranges of $\log_{10}[M_{\rm 200b}/(\hmsun)]\in(13,~13.5],~(14,~14.5]$ and $(15,~15.5]$.
        Each line shows the results for $\gs=0$ and $\pm0.1$. Right: Same as the left panel but for the PDFs of $\Az$, $P(\Az)$.}
        \label{fig:A_mass}
\end{figure*}

Fig.~\ref{fig:A_mass} shows the mass dependence of the modification of the PDFs of $\Ax$ and $\Az$ due to the SA, respectively. From top to bottom, we show the results for the halo mass ranges $\log_{10}[M_{\rm 200b}/(\hmsun)] \in (13,~13.5]$,~$(14,~14.5]$, and $(15,~15.5]$.
The correspondence between line type and $\gs$ values is the same as in Fig.~\ref{fig:A_all}.
We omit $P(\Ay)$ here, as it is nearly identical to $P(\Ax)$ due to the symmetry with respect to $\hat{\mathbf{d}}$.
For $\gs = \pm 0.1$ which are comparable to the upper limit currently obtained from the galaxy clusteing measurements~\cite{sugiyama18}, $P(\Ax)$ and $P(\Az)$ show negligible deviations from the isotropic case in the lowest mass range, while the deviations become more pronounced in the $(14,~14.5]$ bin and even stronger in $(15,~15.5]$, maintaining the same $\gs$-dependence as in Fig.~\ref{fig:A_all}.
The errors are larger in the halo mass range of 
$\log_{10}[M_{\rm 200b}/(\hmsun)]  \in(15,~15.5]$ because the number of halos for this mass range is approximately ten times smaller than that for 
$\log_{10}[M_{\rm 200b}/(\hmsun)]  \in(14,~14.5]$.
These results suggest a stronger SA-induced alignment for more massive halos.
This trend is physically reasonable, as more massive halos have larger volumes roughly proportional to their mass and can more effectively ``feel'' global effects such as SA.

\begin{figure*}[htbp]
\centering
\includegraphics[width=0.5\columnwidth]{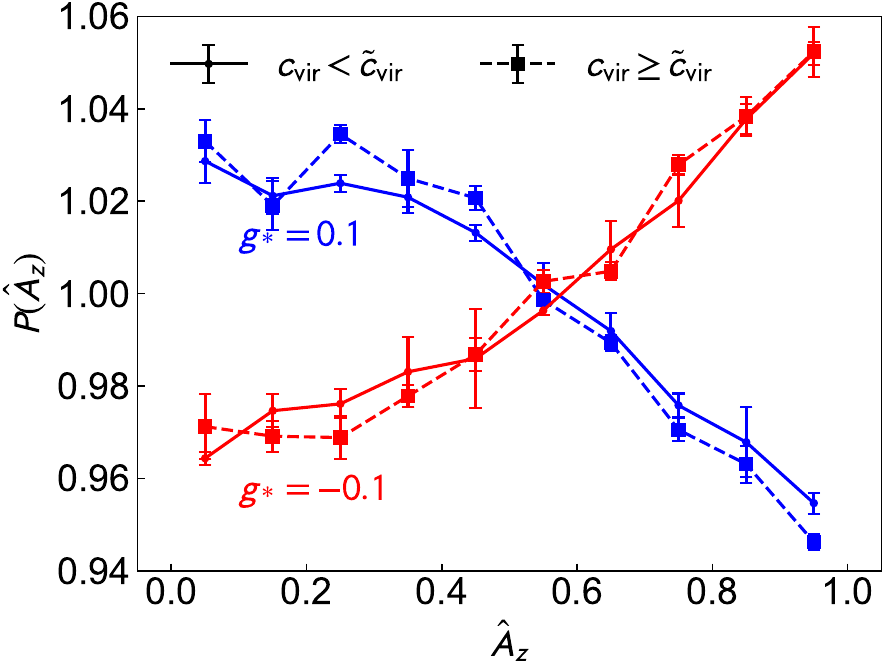}
\caption{PDFs of $\hat A_z$ for the high- and low-$\cvir$ halos of $14<\log_{10}[M_{\rm 200b}/(\hmsun)]\leq14.5$ from the L2 run with $\gs=\pm0.1$.}
\label{fig:P_Az_con}
\end{figure*}
We also examine whether additional factors influence the alignment of halos induced by the SA. In particular, we investigate the correlation between the halo formation epoch and its alignment. It is well established that the halo concentration parameter serves as a reliable proxy for the formation epoch \cite{Wechsler02}.
We define the halo concentration parameter as
\begin{align} c_{\rm vir}=R_{\rm vir}/r_{\rm s}~, \end{align}
where
$R_{\rm vir}$ and $r_{\rm s}$
are, respectively, the virial radius of the halo and the scale radius of the NFW density profile, $\rho_{\rm NFW}(r)=\rho_{\rm s}/[(r/r_{\rm s})(1+r/r_{\rm s})^2]$ \cite{NFW}.
We adopt the values of $R_{\rm vir}$ and $r_{\rm s}$ obtained from the {\sc Rockstar} halo finder.
We divide the halos of $14<\log_{10}[M_{\rm 200b}/(\hmsun)]\leq14.5$ from the L2 run with $\gs=\pm0.1$ into two groups based on $\cvir$: those with concentrations larger and smaller than the median value ${\tilde c}_{\rm vir}$.
Fig.~\ref{fig:P_Az_con} shows the distribution function of $\hat A_z$ for the halos with high and low $\cvir$.
For both values of $\gs$, halos with higher and lower concentrations exhibit similar alignment distributions, indicating that the formation epoch does not significantly affect the halo alignment associated with the SA.
We found the same trend when the halos were divided according to the triaxiality parameter $T$ as well as for halos with masses in the range  $15<\log_{10}[M_{\rm 200b}/(\hmsun)] \leq 15.5$ in the $\gs=\pm0.1$ case.
These results imply that the halo alignment mainly reflects the SA-imprinted initial conditions, rather than the subsequent formation process.

\begin{figure*}
  \centering
  \begin{minipage}{0.45\textwidth}
    \centering
    \includegraphics[width=1.05\columnwidth]{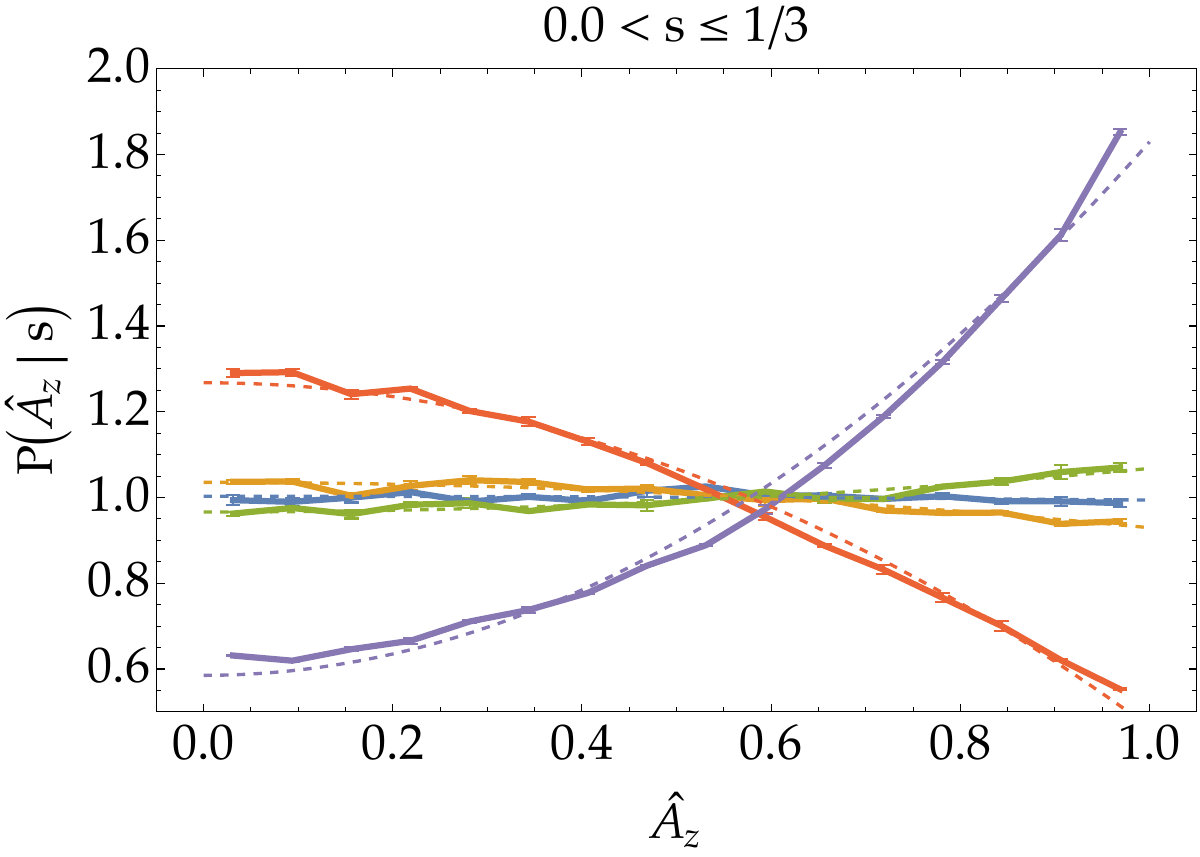}
  \end{minipage}
  \hspace{4mm}
  \begin{minipage}{0.45\textwidth}
    \centering
    \includegraphics[width=1.05\columnwidth]{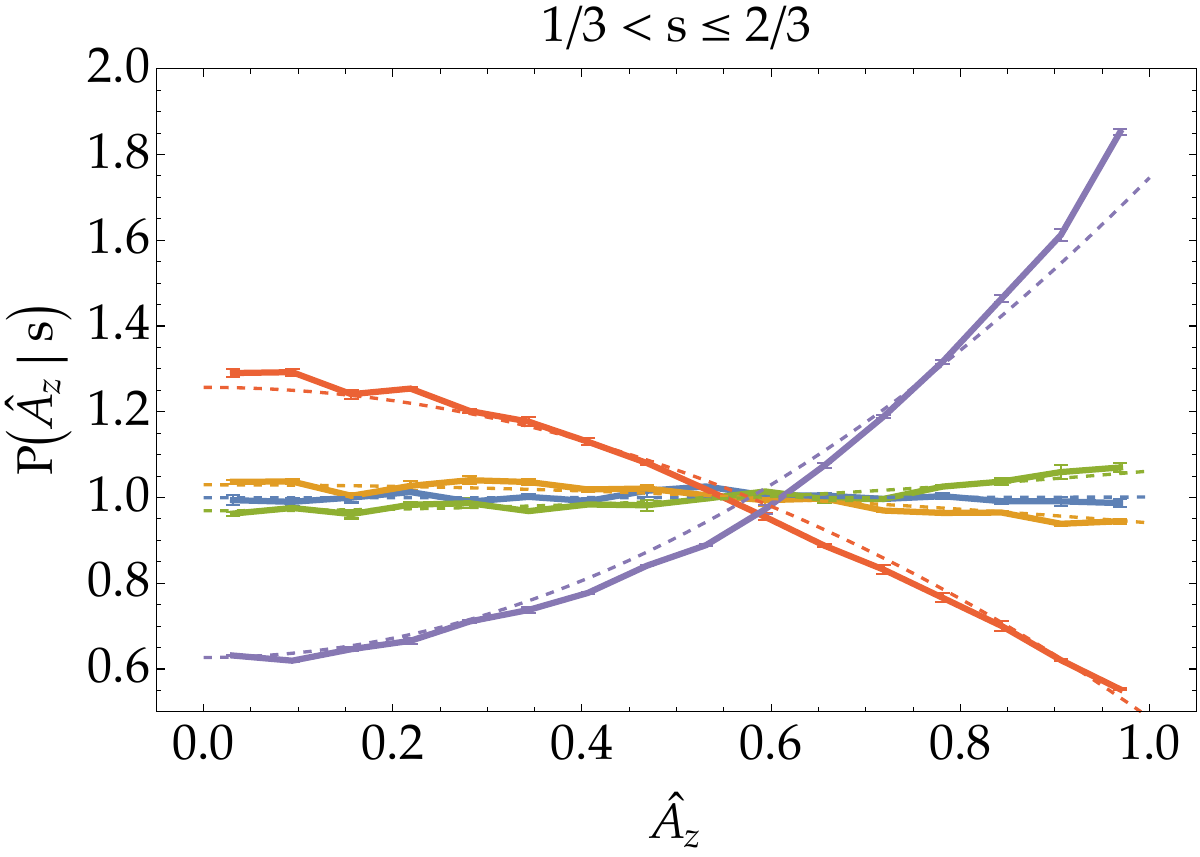}
  \end{minipage}
  \vspace{3mm}
  
  \begin{minipage}{0.4225\textwidth}
    \centering
    \includegraphics[width=1.1\columnwidth]{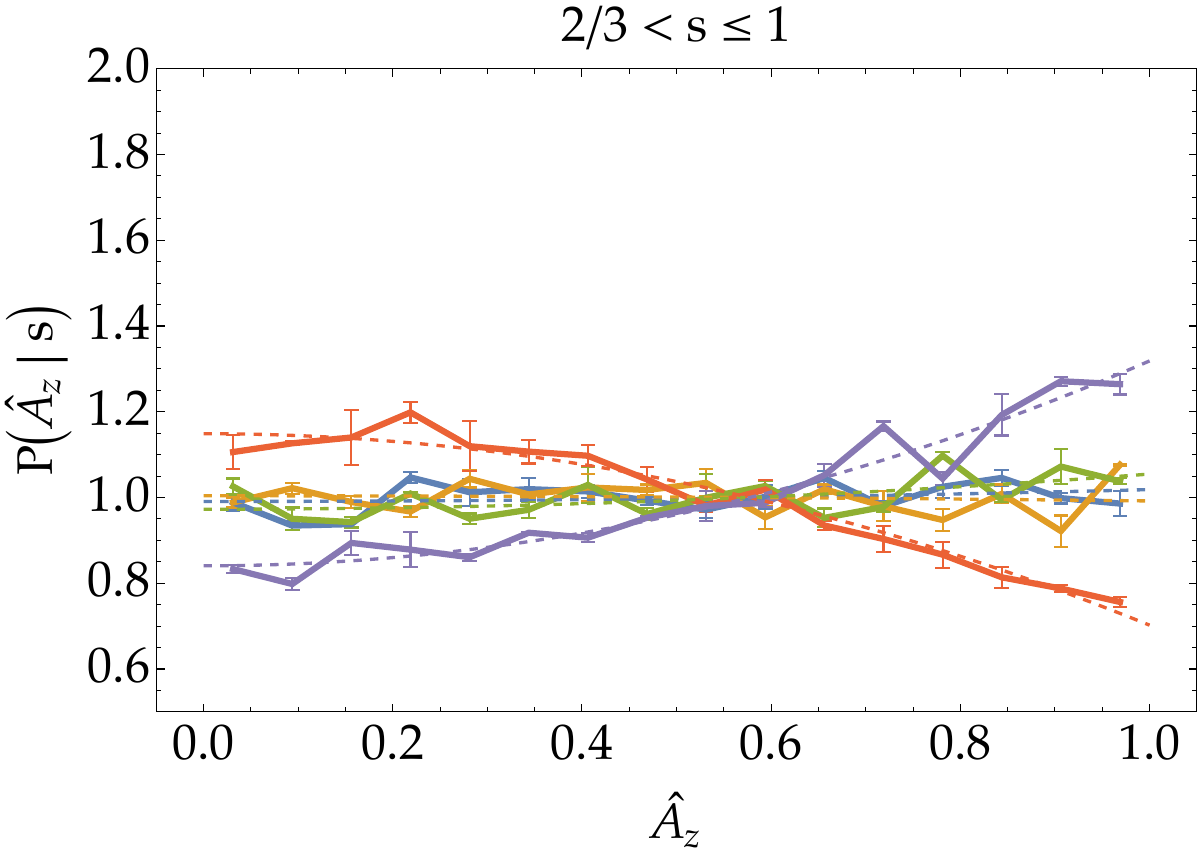}
  \end{minipage}
  \hspace{15mm}
  \begin{minipage}{0.4\textwidth}
    \centering
    \includegraphics[width=1.05\columnwidth]{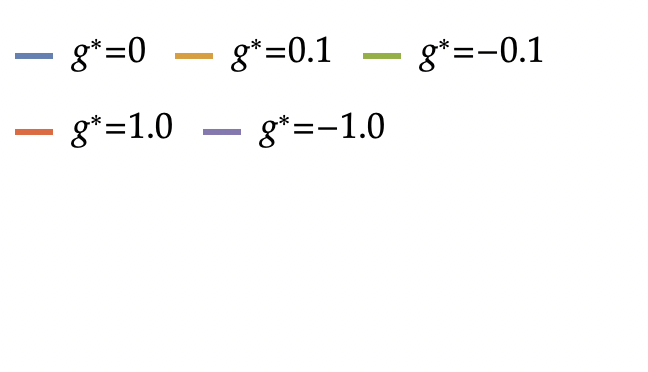}
  \end{minipage}
  \caption{Conditional PDFs of $\Az$, $P(\Az|s)$, for the halo mass ranges of $\log_{10}[M_{\rm 200b}/(\hmsun)]\in~(14,~14.5]$ for each bin of $\Az$, respectively. Each panel displays the results for different range of $s$ value: $0.0<s\le 1/3$ (Top left), $1/3<s\le 2/3$ (Top right) and $2/3<s\le 1$ (bottom left). The colored solid curves and dashed curves correnspond to the results and the second-order Legendre polynomial $\calL_ 2(\Az)$ fit, $1+a \calL_ 2(\Az)$ with fitting parameter $a$, for different $\gs$ values:~$\gs=0$ (blue),~$0.1$ (orange),~$-0.1$ (green),~$1.0$ (red) and $-1.0$ (purple).}
  \label{fig:Conditional_PDF_Az_s}
\end{figure*}
As discussed in Sec.~\ref{the halo shapes}, the remarkable 
insensitivity to $\gs$ of $P(s)$ (Fig.~\ref{fig:ST_prob}) fundamentally arises from the integration over $\Az$. To demonstrate this, we examine the conditional distributions $P(\Az|s)$ by binning the halos into three intervals of $s$ with a width of $\Delta s = 1/3$.
As shown in Fig.~\ref{fig:Conditional_PDF_Az_s}, we found that the conditional distributions of $\Az$ are well-approximated by using the second-order Legendre polynomial $\calL_2(\Az)$, i.e., $P(\Az|s) = 1 + a \mathcal{L}_{2}(\mu)$ with $a$ being a fitting parameter, which would be related to the response of $g_{\ast}$ to the probability distribution. We adopt this functional form, motivated by the fact that the anisotropy considered here corresponds to a quadrupolar SA.
Since the integral of $\calL _2 (\Az)$ over the full range of orientation 
$0 \leq \Az \leq 1$
vanishes (or contributes negligibly to the total magnitude), the $\gs$-dependent features observed in the conditional distributions do not manifest in the marginalized distribution $P(s)$ with respect to orientation. 
It follows that the effect of $\gs$ 
cancels out.

%%%%%%%%%%
\subsection{Other vectors associated with the dynamics of halos}
%%%%%%%%%%
In this section, we extend our investigation to additional vector quantities associated with the dynamics of halos: bulk velocities and angular momenta, in order to examine whether there exist vector quantities with SA-dependent directionality besides the orientation of the halo as an ellipsoid.
We are also motivated to investigate these quantities because SA-distorted matter distributions can affect the matter and halo bulk velocity fields.

%%%%%
\subsubsection{Halo bulk velocities}
%%%%%

\begin{figure*}
  \centering
  \begin{minipage}{0.45\textwidth}
    \centering
    \includegraphics[width=1.05\columnwidth]{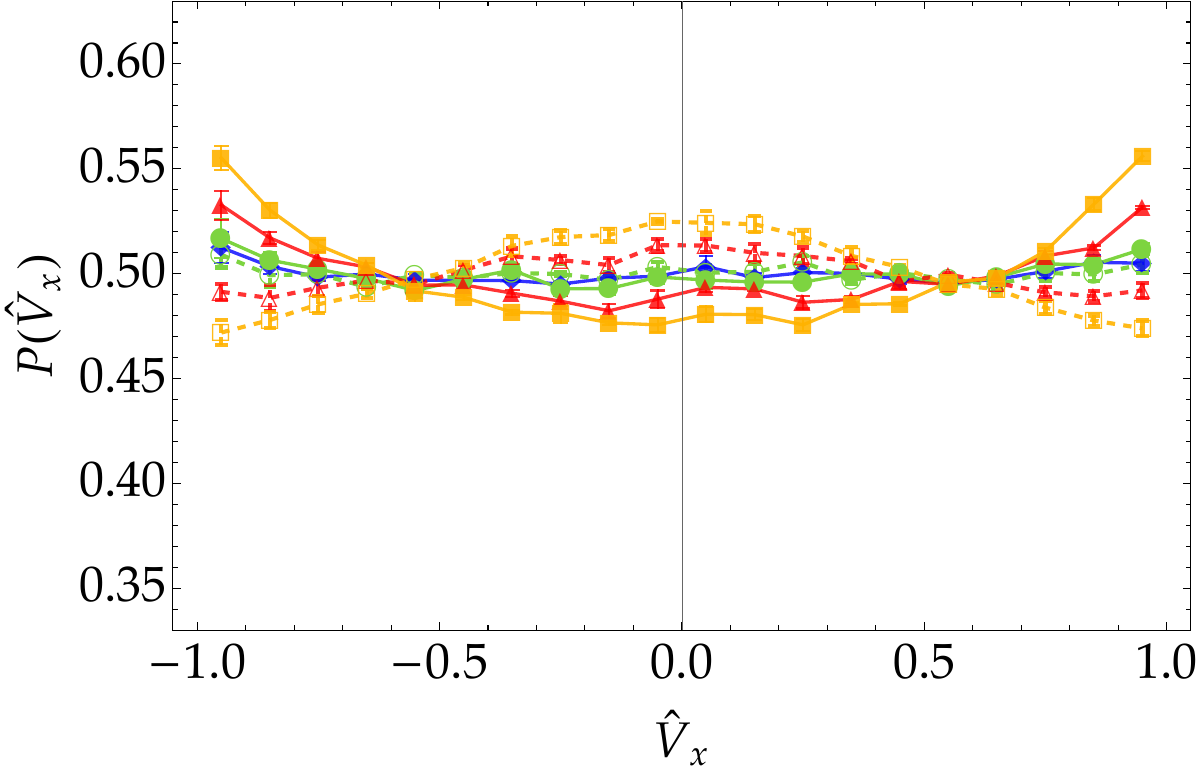}
  \end{minipage}
  \hspace{4mm}
  \begin{minipage}{0.45\textwidth}
    \centering
    \includegraphics[width=1.05\columnwidth]{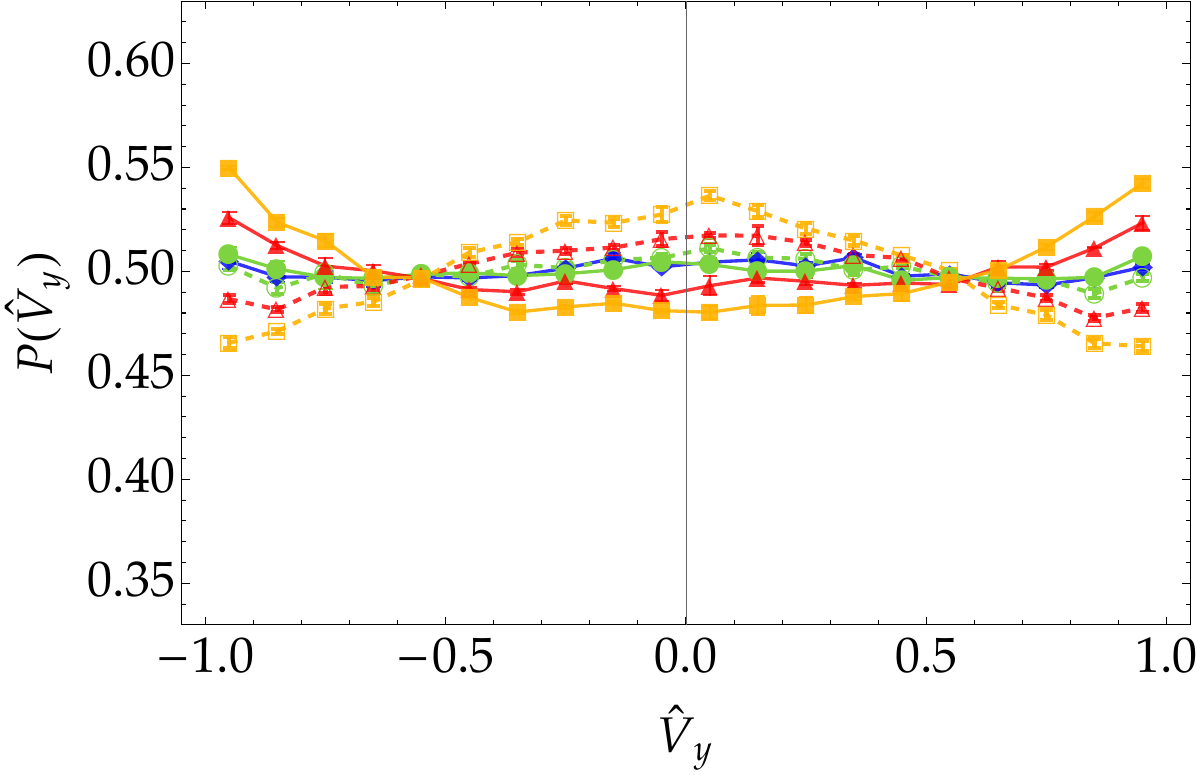}
  \end{minipage}
  \vspace{3mm}
  
  \begin{minipage}{0.4225\textwidth}
    \centering
    \includegraphics[width=1.1\columnwidth]{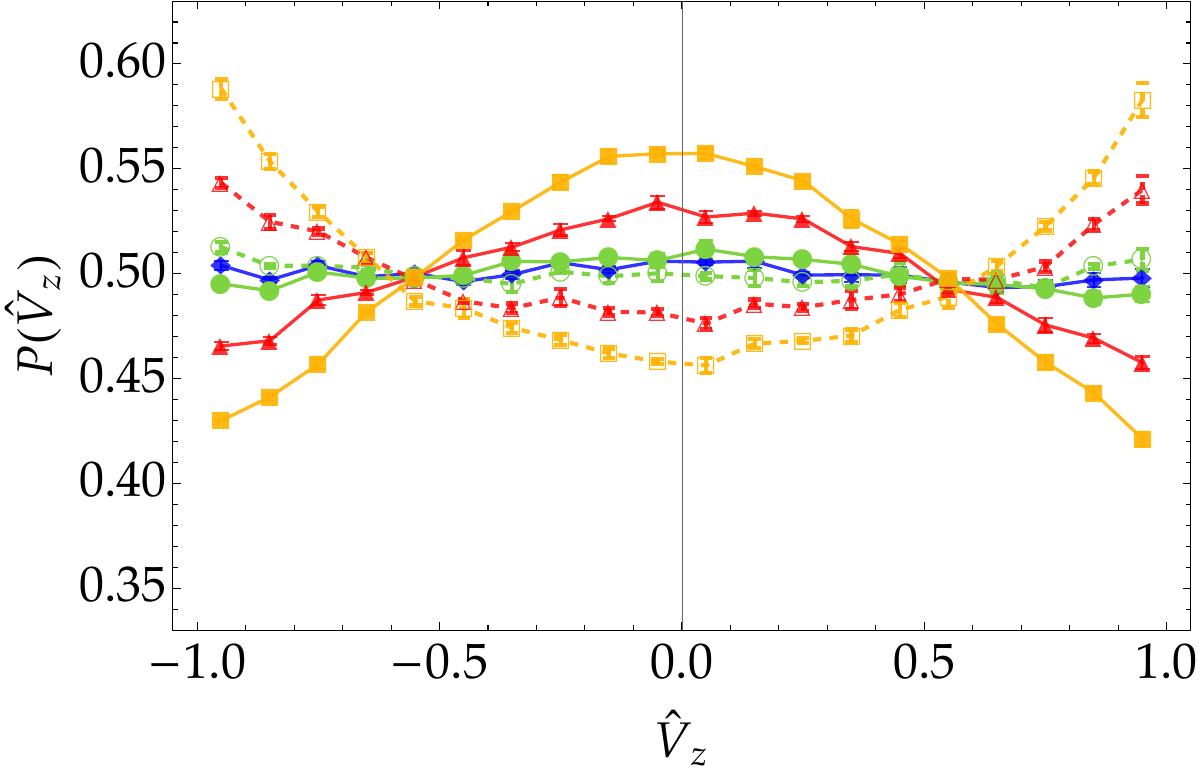}
  \end{minipage}
  \hspace{15mm}
  \begin{minipage}{0.4\textwidth}
    \centering
    \includegraphics[width=1.05\columnwidth]{Figure/Legends.pdf}
  \end{minipage}
  \caption{PDFs of the halo bulk velocities $\Vx$ (left top), $\Vy$ (right top), and $\Vz$ (left bottom), for the halos with masses of $14<\log_{10}[M_{\rm 200b}/(\hmsun)]\leq14.5$ from the L2 run.
  Each color line shows the result for $\gs =0$ (blue),~$\pm 0.1$ (green),~$\pm 0.5$ (red) and $\pm 1.0$ (orange). The dashed (solid) lines stand for the positive (negative) sign.}
  \label{fig:V_All}
\end{figure*}

\begin{figure*}
    \centering
  \begin{minipage}{0.45\textwidth}
    \centering
    \includegraphics[width=1.05\columnwidth]{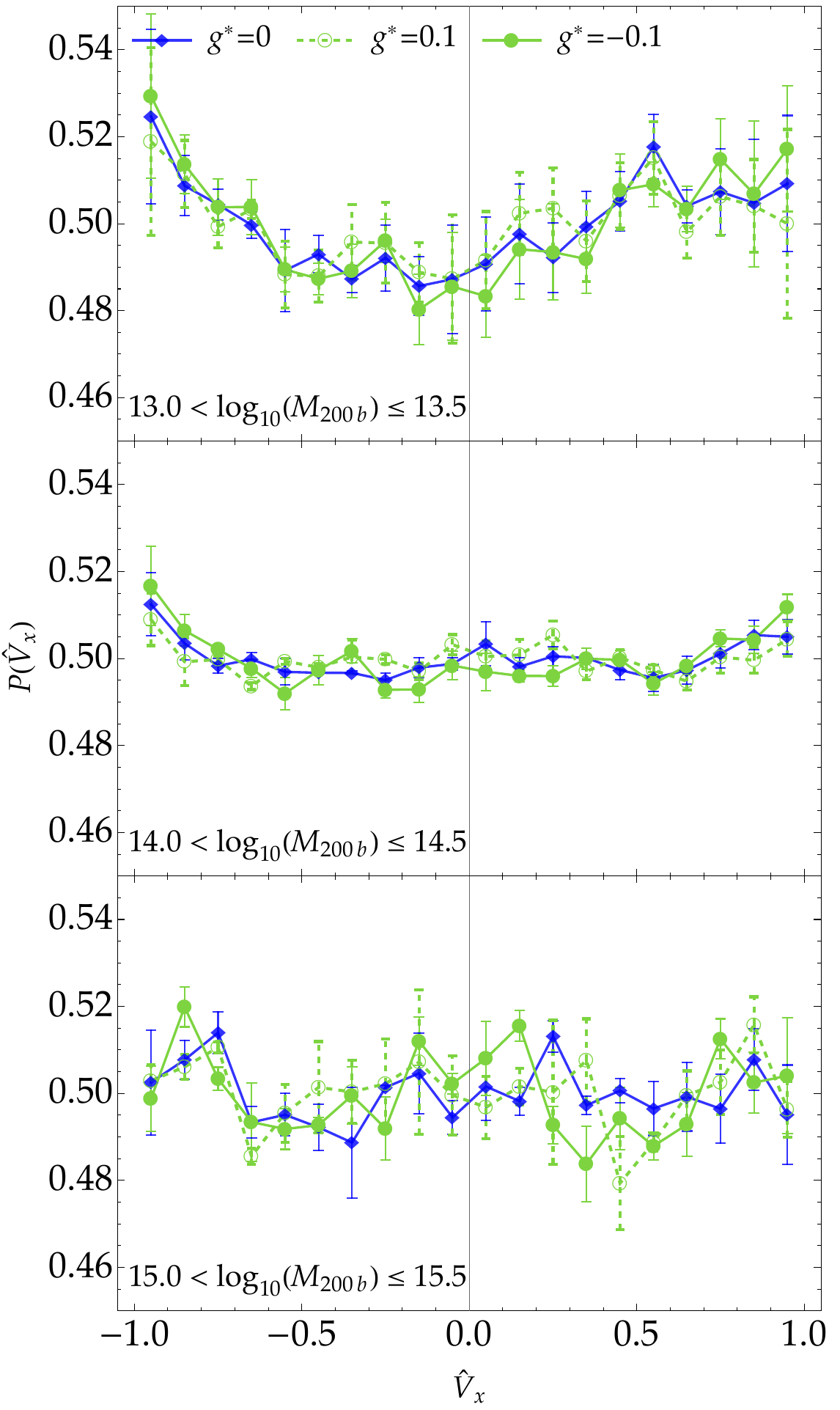}
  \end{minipage}
  \hspace{4mm}
    \begin{minipage}{0.45\textwidth}
    \centering
    \includegraphics[width=1.05\columnwidth]{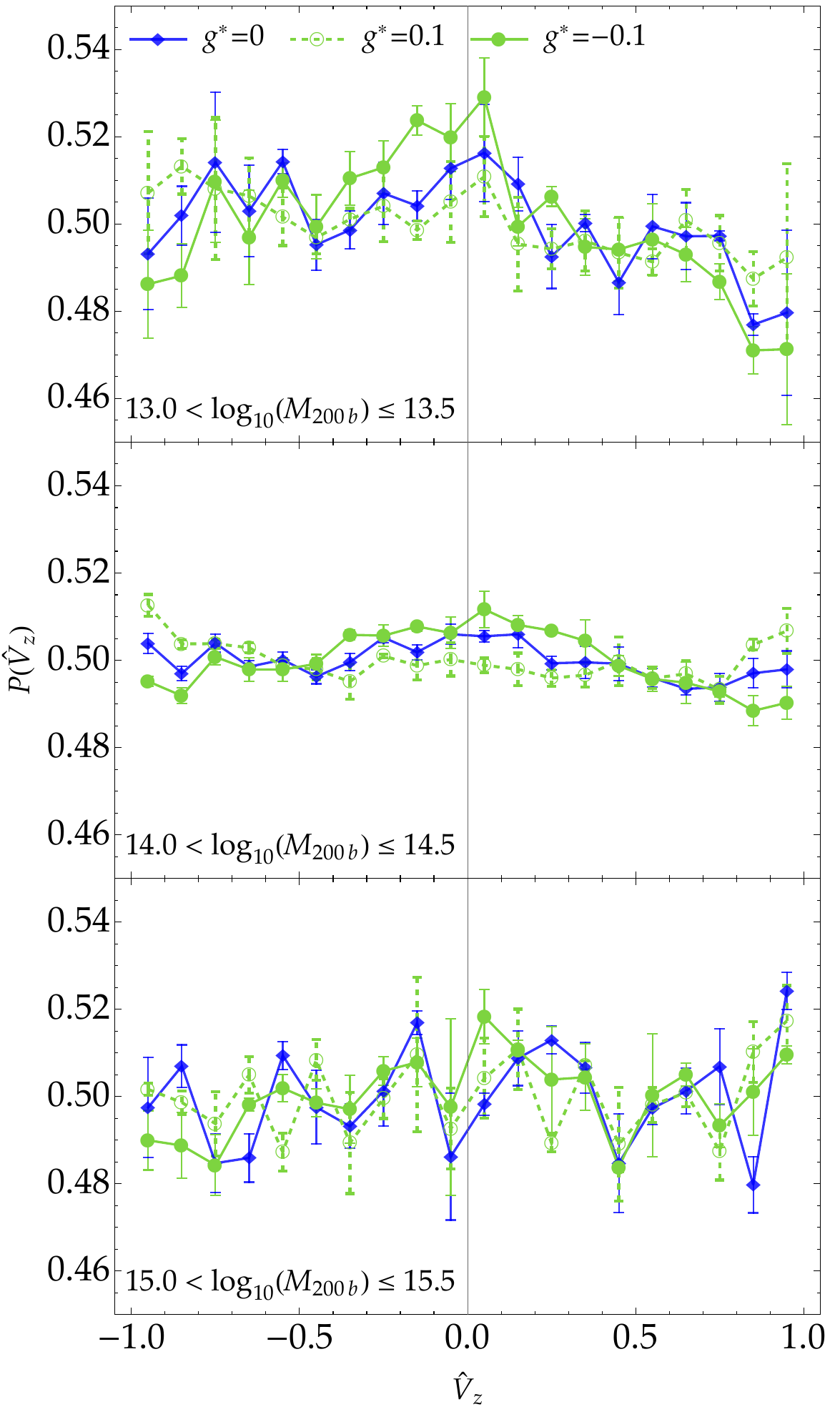}
    \end{minipage}
    \caption{Left: PDFs of $\Vx$, $P(\Vx)$, for the halo mass ranges of $\log_{10}[M_{\rm 200b}/(\hmsun)]\in(13,~13.5],~(14,~14.5]$ and $(15,~15.5]$.
        Each line is the result for $\gs=0$ and $\pm0.1$. Right: Same as the left panel but for the PDFs of $\Vz$, $P(\Vz)$.}
    \label{fig:V_mass}
\end{figure*}

To examine whether the halo bulk velocity vector, ${\bf{V}}=(V_x, ~V_y, ~V_z)$, carries any imprint of SA, we focus on the unit vector of the bulk velocity of a halo $\hat{{\bf V}} = {\bf V}/|{\bf V}|$, and study the PDFs of each component for several values of $\gs$.
Fig.~\ref{fig:V_All} shows the PDFs of the bulk velocity component within the mass range 
$\log_{10}[M_{\rm 200b}/(\hmsun)]  \in(14,~14.5]$.
In this figure, we observe a similar trend to that seen in the PDFs of halo orientations in Fig.~\ref{fig:A_all}, indicating that the bulk velocity of the halo is also affected by the SA. Note that the dependence on the sign of $\gs$ of the bulk velocity vector distributions is observed to be opposite to that of the orientations.
In the case of $g_{*} < 0$ ($g_{*} > 0$), structures form preferentially parallel (perpendicular) to the $z$-direction, and the associated gravitational potential wells are also aligned along the parallel (perpendicular) directions to the $z$-axis. Consequently, halos are more likely to fall into these potential wells, which increases the number of bulk-velocity vectors aligned perpendicular (parallel) to the $z$-axis.
However, it is clear that sensitivity to the SA of the bulk velocity vectors is still weaker compared to the orientations. This is more clearly shown in Fig.~\ref{fig:V_mass}.
Fig.~\ref{fig:V_mass} illustrates how the effect of SA on the bulk velocity depends on halo mass for $\gs = \pm 0.1$. 
We find that the bulk velocity vectors do not significantly depend on the SA with respect to the size of the statistical error.
We note that the probability distributions $P({\Vx})$ for the mass range of 
$\log_{10}[M_{\rm 200b}/(\hmsun)]\in(13,~13.5]$, and $\in(14,~14.5]$ exhibit a mildly downward convex profile around $\Vx=0.0$ even for the case with $\gs = 0$. 
Further data and realizations will be required to confirm this feature with higher statistical significance.
%

%%%%%
\subsubsection{Halo angular momenta}\label{angular}
%%%%%

\begin{figure*}
  \centering
  \begin{minipage}{0.45\textwidth}
    \centering
    \includegraphics[width=1.05\columnwidth]{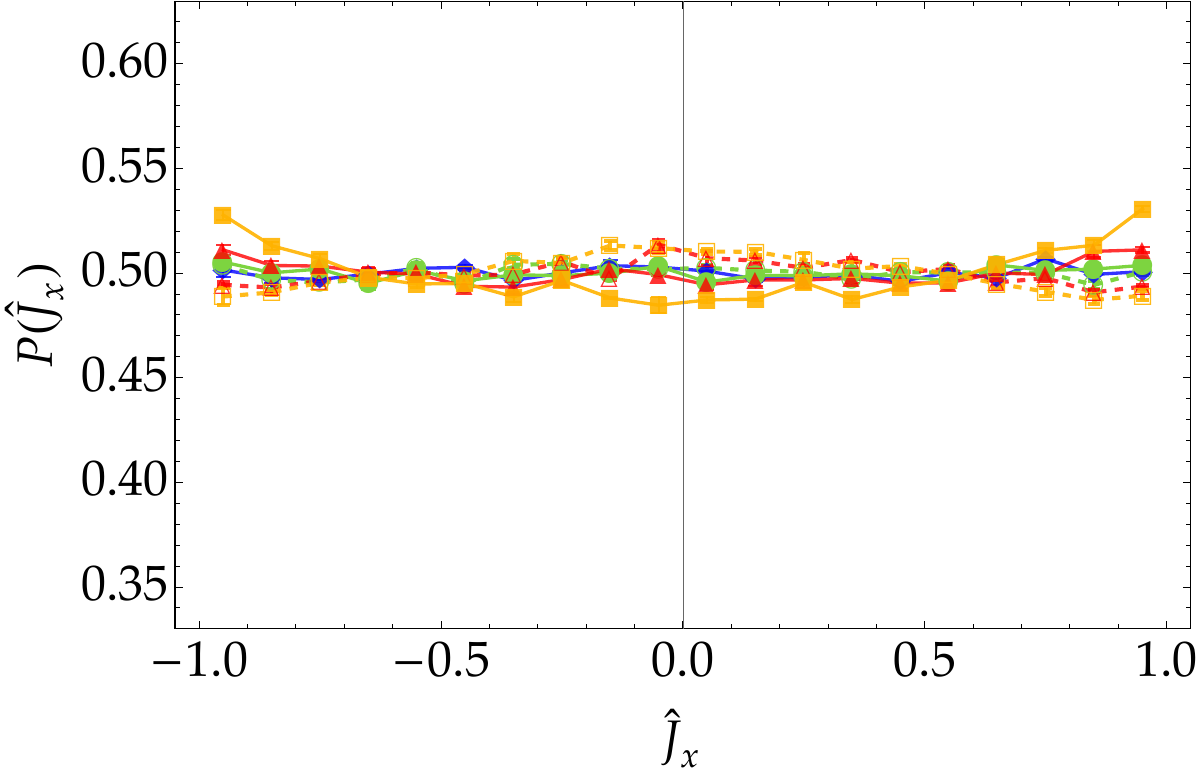}
  \end{minipage}
  \hspace{4mm}
  \begin{minipage}{0.45\textwidth}
    \centering
    \includegraphics[width=1.05\columnwidth]{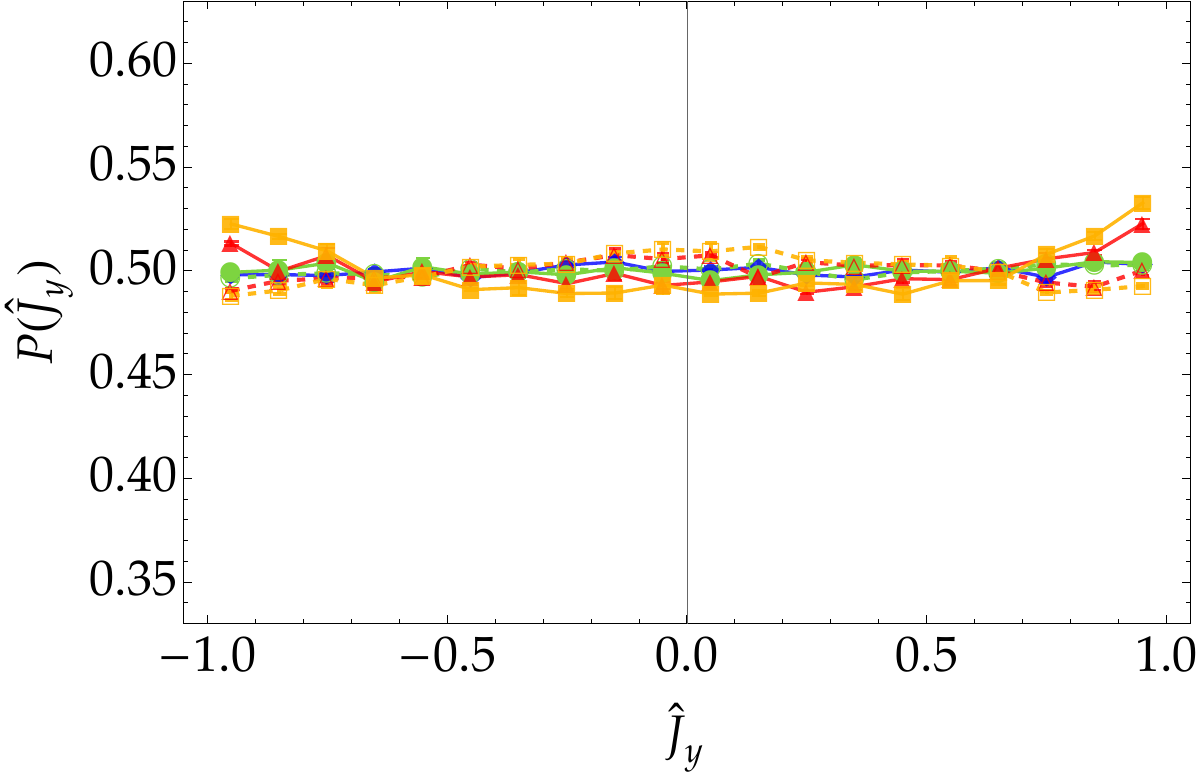}
  \end{minipage}
  \vspace{3mm}
  
  \begin{minipage}{0.4225\textwidth}
    \centering
    \includegraphics[width=1.1\columnwidth]{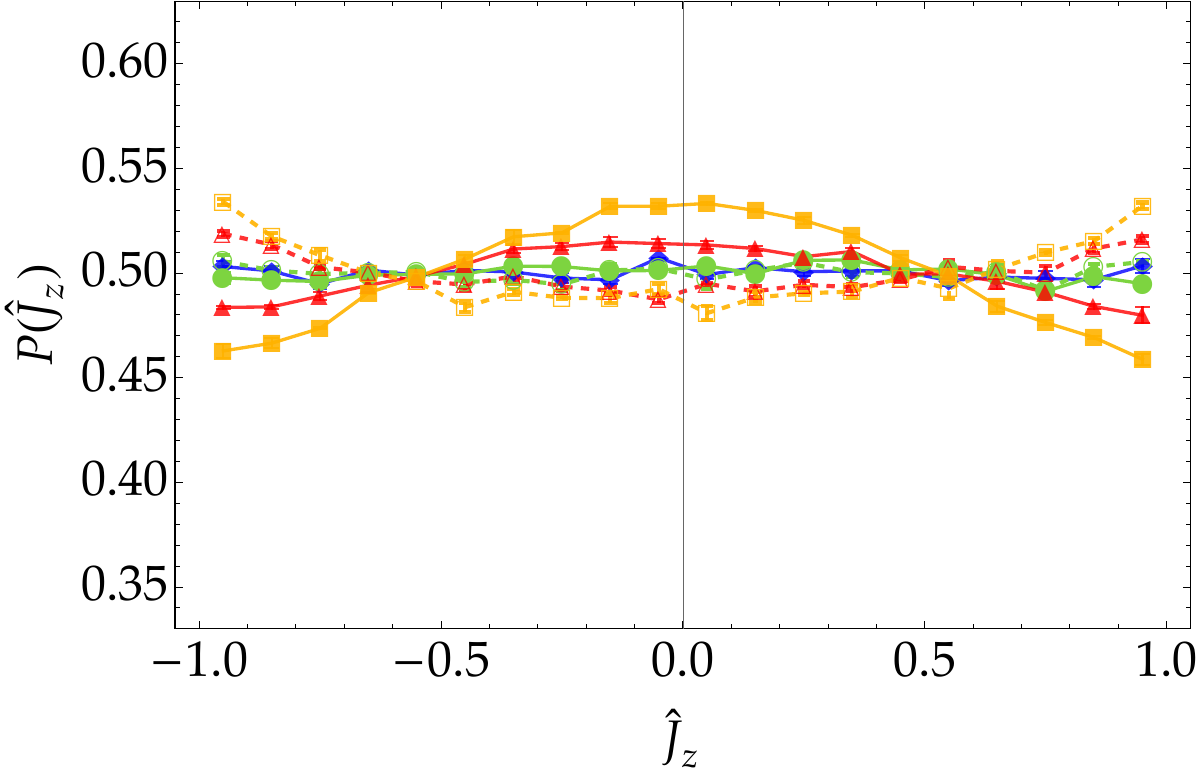}
  \end{minipage}
  \hspace{15mm}
  \begin{minipage}{0.4\textwidth}
    \centering
    \includegraphics[width=1.05\columnwidth]{Figure/Legends.pdf}
  \end{minipage}
  \caption{PDFs of $\Jx,~\Jy$ and $\Jz$, i.e., the halo angular momenta vector, for the halos with masses of $14<\log_{10}[M_{\rm 200b}/(\hmsun)]\leq14.5$ from the L2 run.
  Each color line shows the result for $\gs =0$ (blue),~$\pm 0.1$ (green),~$\pm 0.5$ (red), and $\pm 1.0$ (orange), and the dashed (solid) lines are for the cases with the positive (negative) sign.}
  \label{fig:J_All}
\end{figure*}

\begin{figure*}
    \centering
  \begin{minipage}{0.45\textwidth}
    \centering
    \includegraphics[width=1.05\columnwidth]{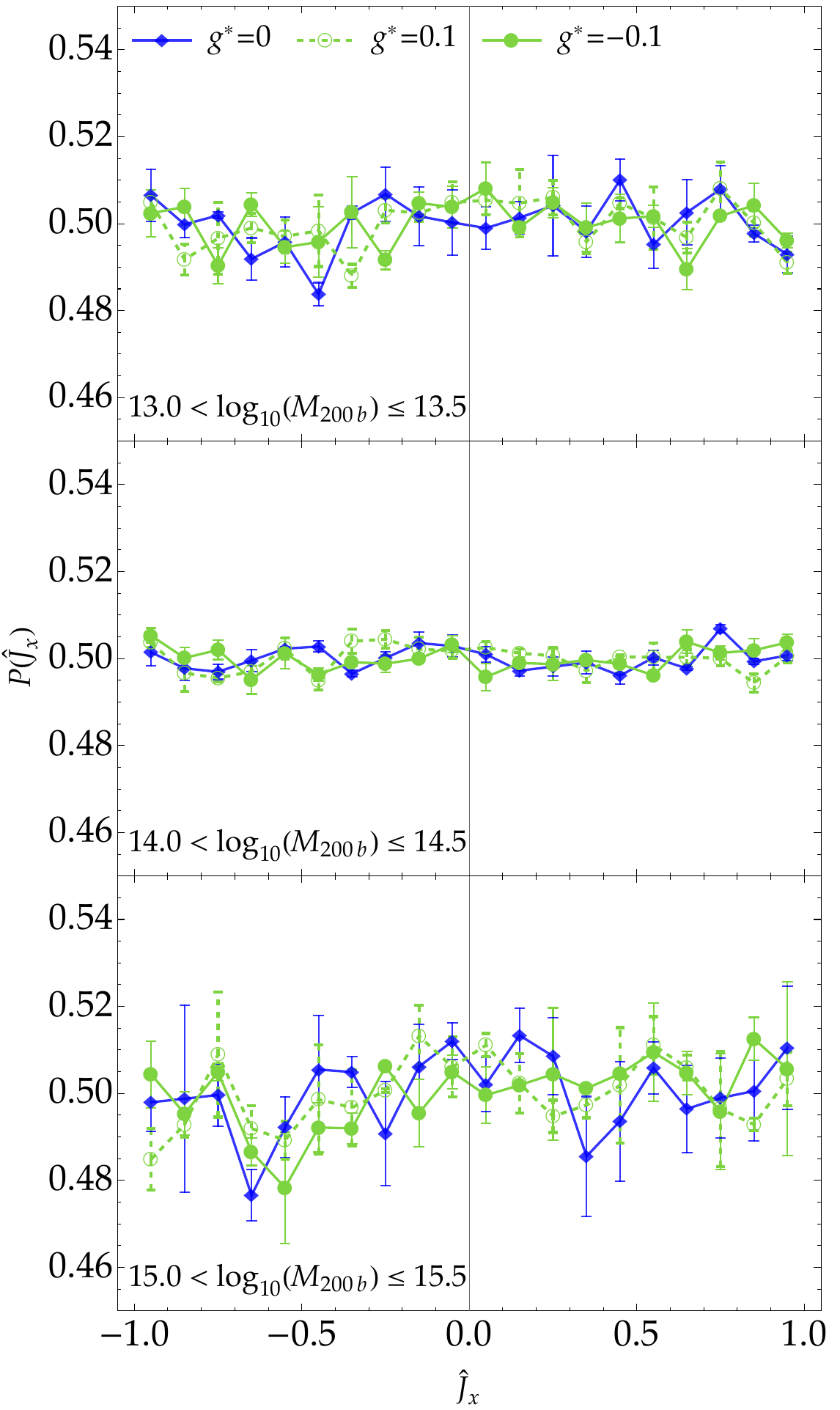}
  \end{minipage}
  \hspace{4mm}
  \begin{minipage}{0.45\textwidth}
    \centering
    \includegraphics[width=1.05\columnwidth]{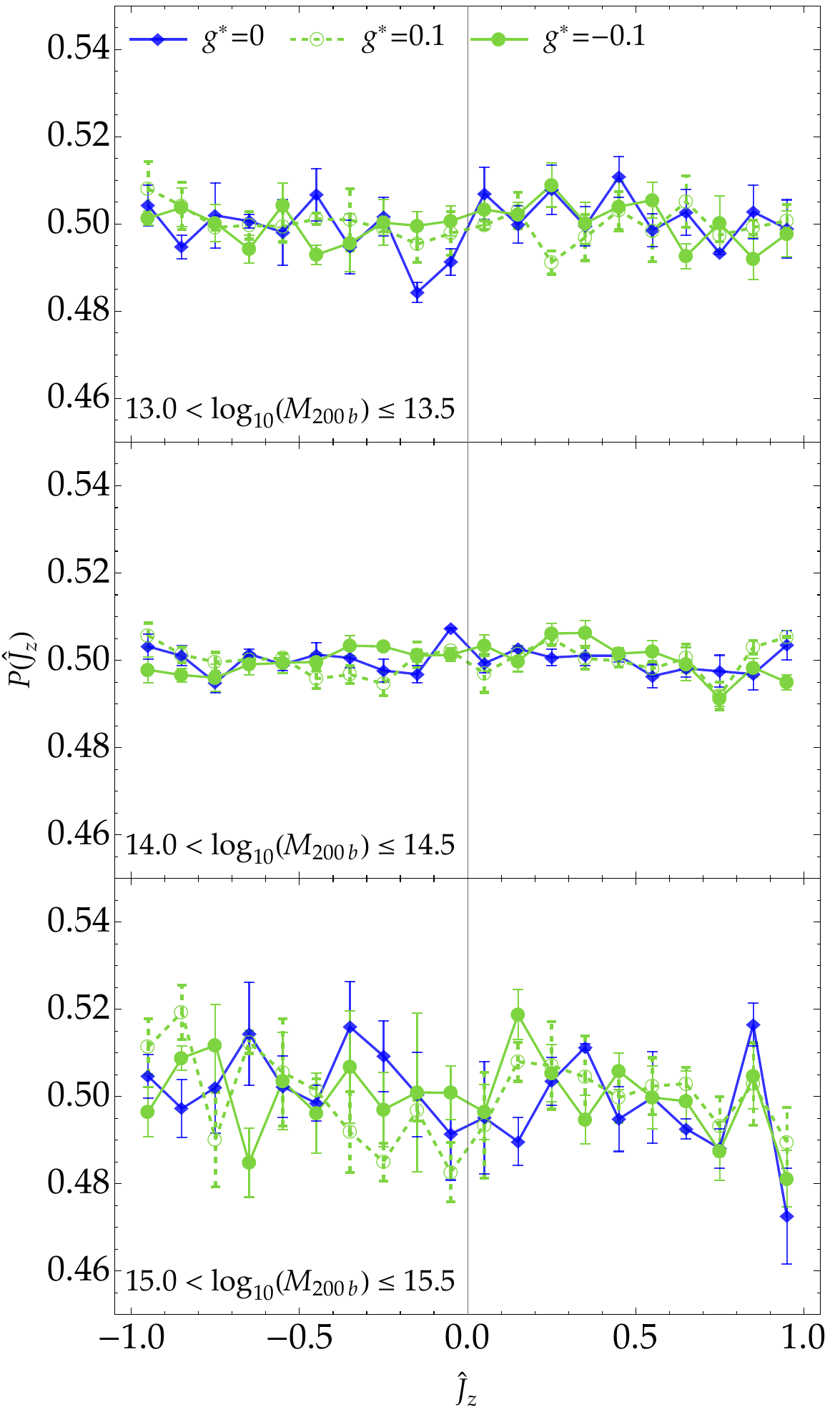}
  \end{minipage}
  \caption{Left: PDFs of $\Jx$, $P(\Jx)$, for the halo mass ranges of $\log_{10}[M_{\rm 200b}/(\hmsun)]\in(13,~13.5],~(14,~14.5]$ and $(15,~15.5]$.
        Each line is the result for $\gs=0$ and $\pm0.1$. Right: Same as the left panel but for the PDFs of $\Jz$, $P(\Jz)$.}
    \label{fig:J_mass}
\end{figure*}

We also study the angular momentum vector of each halo provided in the 
{\sc Rockstar} halo catalog, ${\bf{J}}=(J_x, ~J_y, ~J_z)$ in the 
SA universe. 
To highlight the directional information of the angular momentum without being dominated by its amplitude, we focus on the unit vector of the angular momenta: $\hat{\bf J}={\bf J}/|{\bf J}|$.
In Fig. \ref{fig:J_All}, we present the distributions of the components of $\hat{\bf J}$ for halos in the mass range $\log_{10}[M_{\rm 200b}/(\hmsun)]\in(14,~14.5]$.
Each panel includes results for various values of the SA parameter $\gs=0,~\pm0.1,~\pm0.5,~\pm1.0$, allowing us to visually inspect potential trends. 
In this figure, a similar trend to that seen in the PDFs of halo bulk velocities in Fig.~\ref{fig:V_All} can be observed, indicating that the direction of the angular momentum of the halo is also affected by the SA. Note that the dependence on the sign of $\gs$ of the angular momentum distributions is observed to be opposite to that of the orientations. 
This behavior originates from the trend for the angular momentum of halos to be oriented parallel to its minor-axis vector as already found in Refs.~\cite{1992ApJ...399..405W,Bailin:2004wu,Shaw:2005dy,Bett:2006zy,2006MNRAS.367.1781A}, though our results demonstrate that this characteristic persists even in the presence of SA.
However, it is clear that sensitivity to the SA of the angular momentum is much weaker compared to the orientations and bulk velocities.
Fig.~\ref{fig:J_mass} shows the halo-mass dependence of the impact of SA on the halo angular momenta $\hat{J}_x$ and $\hat{J}_z$ for $g_* = \pm 0.1$.
The comparison across mass ranges reveals that there is no statistically significant dependence of the SA effect on the angular momentum for the $g_*$ that is comparable to the upper limit from the observations of galaxy clustering \cite{sugiyama18}. 
Consequently, the angular momentum vector is less sensitive than the orientation and bulk velocity for probing the SA effect.

When taking into account previous sections and Fig.~\ref{fig:A_mass} (orientations) together with the results from Fig.~\ref{fig:V_mass} (bulk velocities) and Fig.~\ref{fig:J_mass}  (angular momenta), it becomes evident that orientation is the most sensitive probe of SA among the quantities in this work, examined for $\gs=\pm 0.1$ which are comparable to the upper limit currently obtained from the galaxy clusteing measurements~\cite{sugiyama18}.

%%%%%%%%%%%%%%%%%%%
\section{Conclusion}\label{conclusion}
%%%%%%%%%%%%%%%%%%%
In this work, we investigated the impact of the SA in the matter density field on the shapes and orientations of cluster-sized halos at $z=0$ using cosmological $N$-body simulations that incorporate a quadrupolar-type SA characterized by the magnitude parameter $\gs$.
For halo shapes, we examined the three-dimensional axis ratio $s$ (smallest-to-largest) and the triaxiality parameter $T$.
Halo orientations were represented by the direction of the largest axis as an ellipsoid, denoted by the vector $\mathbf{\hat A}$.
 We found that the PDFs of both of the shape parameters, $s$ and $T$, exhibit little sensitivity to SA (Fig.~\ref{fig:ST_prob}). This is because the $\gs$-dependence effectively cancels out upon 
marginalization, a consequence of the specific symmetry of the conditional PDF (Fig.~\ref{fig:Conditional_PDF_Az_s}). In contrast, %while both of the shape parameters, $s$ and $T$, show little sensitivity to SA (Fig.~\ref{fig:ST_prob}), 
the halo orientations are significantly affected (Fig.~\ref{fig:A_all}).
We also found that the halo orientations are significantly affected (Fig.~\ref{fig:A_all}).
We also found that halo orientations tend to align perpendicular to the SA-preferred direction $\hat{\mathbf{d}}$ for positive values of $\gs$, and parallel for negative values (Fig.~\ref{fig:A_all}).

We used halos with masses of $\log_{10}[M_{\rm 200b}/(\hmsun)] \in (13,~13.5]$,~$(14,~14.5]$, and $(15,~15.5]$, and investigated the mass dependence of the alignment due to the SA.
We clearly showed that this SA alignment effect becomes more pronounced for more massive halos, suggesting that more massive halos are more sensitive to the global anisotropic nature of the matter distribution (Fig.~\ref{fig:A_mass}).
Importantly, this alignment is expected to originate from the initial conditions that have inherent information about the SA, rather than later-stage nonlinear processes, as we found no significant dependence on halo properties such as concentration or ellipticity (Fig.~\ref{fig:P_Az_con}). We also investigated the sensitivity of other vector quantities related to the dynamics of halos, such as the halo bulk velocity and angular momentum, to the SA effect. 
Although their PDFs show that these quantities are less sensitive to SA than halo orientations, the impact on bulk velocity is particularly noteworthy. Since changes in the velocity field are expected to manifest in redshift-space distortions, a detailed comparison with observational data remains a compelling direction for future work. Overall, our comparison among bulk velocity, angular momentum, and orientation leads us to conclude that halo orientations are the most sensitive to SA among the quantities examined in this study.
%Although their PDFs show qualitatively similar responses to the SA, which is opposite to that of the halo orientations, the magnitude of these changes is smaller than that seen in the halo orientations. 
%This comparison leads us to conclude that the orientations are the most sensitive among the examined quantities in this paper, and thus serve as a promising new probe of SA.

In this paper, we employed a simple scale-independent quadrupole type of SA, but other types of SA are actually possible. In fact, in the context of the inflation models with gauge fields, including the anisotropic inflation, the possibility of generating various types of SA has been actively discussed, e.g., statistically anisotropic non-Gaussianity~(see, e.g., Ref.~\cite{Yokoyama:2008xw}), the SA in the primordial gravitational waves~(see, e.g., Ref.~\cite{Fujita:2018zbr}), and so on. 
In conjunction with those observations, it would be interesting to apply our analysis to more specific model-based types of SA to verify the early universe.

As the PDFs presented in this paper are one-point functions, they are insensitive to spatial correlations in halo orientations.  
However, the SA-induced changes in halo orientations may have implications for other cosmological observables, such as intrinsic alignments (IA) \cite{croft00,heavens00,lee00} (see also Refs.~\cite{shi24,ishikawa25} for recent studies on the IA of galaxy clusters).  
Since these effects could be potential sources of systematic errors in cosmological analyses or serve as tools to constrain the SA parameter, investigating how SA influences a broader range of statistical measures is an important direction for future work.

Our findings highlight the potential of using halo alignment as a complementary probe of the SA in the Universe.
In particular, measurements of projected halo ellipticity, which is observable in galaxy cluster-galaxy lensing observations, would exhibit systematic directional features due to the SA-induced alignment and could be used to constrain $\gs$.
However the expected number of galaxy clusters with masses around $10^{14}M_\odot$ at $z \simeq 0$ in the photometric {\it Euclid} survey is $\sim 10^4$ \cite{2016MNRAS.459.1764S},while the average number of halos in our study is $\sim 10^5$.
Assuming this sample size and pure Poisson statistics, the error bars on $P(\hat{A}_z)$ shown in Figure~4 would be $\simeq 0.04$, whereas the maximum deviation induced by SA from the isotropic case is only $\simeq 0.05$. Therefore, even if halo orientation were a directly observable quantity, a modern wide-area survey such as {\it Euclid} alone would be unlikely to provide a significant detection of the SA signature only from the one-point statistics. A possible way to improve the constraining power would be to combine multiple observables sensitive to halo alignment, for example, the projected halo ellipticity inferred from galaxy--cluster lensing in Ref.~\cite{oguri10} and the spatial distribution of cluster galaxies in Ref.~\cite{ishikawa25}, possibly including their redshift dependence. We leave a more quantitative investigation of these possibilities to future work.
% Further issues, including projection effects, redshift evolution, and connections to observable quantities, will be necessary to assess the feasibility of such constraints in future surveys.

%%%%%%%%%%%%%%%%%%%
\section*{Acknowledgements}
%%%%%%%%%%%%%%%%%%%
We thank Kazuyuki Akitsu, Maresuke Shiraishi, Takahiro Nishimichi, and Teppei Okumura for their useful comments and discussions. We also thank the anonymous referee for providing comments that helped to deepen the discussion.
The simulations were carried out on Cray XC50 and XD2000 at the Center for Computational Astrophysics, National Astronomical Observatory of Japan, and at the Yukawa Institute Computer Facility.
This work is supported by JSPS KAKENHI Grants No. JP22K03644 (S.M.), No. 24K17043 (S.S.), No. JP20K03968, No. JP23H00108, and No. JP24K00627 (S.Y.). Y.M. is supported by JST SPRING, Grant Number JPMJSP2125, and “THERS Make New Standards Program for the Next Generation Researchers.”

\bibliographystyle{JHEP}
\bibliography{lssref.bib}

\end{document}